\documentclass[journal,twocolumn]{IEEEtran}

\ifCLASSINFOpdf
  \usepackage[pdftex]{graphicx}
  \graphicspath{{../pdf/}{../jpeg/}}
  \DeclareGraphicsExtensions{.pdf,.jpeg,.png}
\else
\fi
\usepackage{xcolor}

\usepackage{amsmath}
\usepackage{algorithmic}
\usepackage{array}
\usepackage{multirow}

\ifCLASSOPTIONcompsoc
 \usepackage[caption=false,font=normalsize,labelfont=sf,textfont=sf]{subfig}
\else
  \usepackage[caption=false,font=footnotesize]{subfig}

\usepackage{stfloats}

\usepackage{flushend}

\usepackage{url}
\usepackage{listings}
\usepackage{color}

\definecolor{dkgreen}{rgb}{0,0.6,0}
\definecolor{gray}{rgb}{0.5,0.5,0.5}
\definecolor{mauve}{rgb}{0.58,0,0.82}

\begin{document}

\title{Security Analysis Methods on Ethereum Smart Contract Vulnerabilities --- A Survey}

\author{Purathani Praitheeshan$^{\star}$, Lei Pan$^{\star}$, Jiangshan Yu$^{\dagger}$, Joseph Liu$^{\dagger}$, and Robin Doss$^{\star}$%
\thanks{Corresponding authors: Lei Pan and Jiangshan Yu, email: l.pan@deakin.edu.au and jiangshan.yu@monash.edu}%
\thanks{$^{\star}$ School of Information Technology, Deakin University, Geelong, VIC 3220, Australia}%
\thanks{$^{\dagger}$ Faculty of Information Technology, Monash University, Clayton, VIC Australia}%
\thanks{This work has been submitted to the IEEE for possible publication. Copyright may be transferred without notice, after which this version may no longer be accessible.}}%

\maketitle

\begin{abstract}
Smart contracts are software programs featuring both traditional applications and distributed data storage on blockchains. Ethereum is a prominent blockchain platform with the support of smart contracts. The smart contracts act as autonomous agents in critical decentralized applications and hold a significant amount of cryptocurrency to perform trusted transactions and agreements. Millions of dollars as part of the assets held by the smart contracts were stolen or frozen through the notorious attacks just between 2016 and 2018, such as the DAO attack, Parity Multi-Sig Wallet attack, and the integer underflow/overflow attacks. These attacks were caused by a combination of technical flaws in designing and implementing software codes. However, many more vulnerabilities of less severity are to be discovered because of the scripting natures of the Solidity language and the non-updateable feature of blockchains. Hence, we surveyed 16 security vulnerabilities in smart contract programs, and some vulnerabilities do not have a proper solution. This survey aims to identify the key vulnerabilities in smart contracts on Ethereum in the perspectives of their internal mechanisms and software security vulnerabilities. By correlating 16 Ethereum vulnerabilities and 19 software security issues, we predict that many attacks are yet to be exploited. And we have explored many software tools to detect the security vulnerabilities of smart contracts in terms of static analysis, dynamic analysis, and formal verification. This survey presents the security problems in smart contracts together with the available analysis tools and the detection methods. We also investigated the limitations of the tools or analysis methods with respect to the identified security vulnerabilities of the smart contracts.

%\color{blue} Please start to add the most important information here: 1st sentence --- motivation of the paper, 2nd sentence --- the security problems of smart contracts, 3rd sentence --- the research questions answered by this paper, 4th sentence --- survey method and how literature are collected and analyzed, 5th sentence --- the important and more interesting findings, 6th sentence --- contributions of this survey, 7th sentence --- potential impact of this survey. See \url{https://www.sciencedirect.com/science/article/pii/S0167404817302511} for example. 
\end{abstract}


\section{Introduction}
Traditional financial systems comfort with the centralized environment where a trusted third party manages and validates the transactions from one party to another \cite{RN68, RN100}. Having an intermediary or regulator to process a valuable transaction in a secured platform is essential \cite{RN152}. Though a centralized environment is a reliable and trustworthy method, its drawbacks are manifold: The processing time for transactions may vary from one hour to a few days; the transaction cost charged by the third party service provider, such as banks or non-financial institutions, is an unnecessary expense for the user \cite{RN153}. In consequence of these issues of the traditional financial systems, the technology advances in peer to peer network and decentralized data management were headed up as the way of mitigation. In recent years, the blockchain technology is being the prominent mechanism which uses distributed ledger technology (DLT) to implement digitalized and decentralized public ledger to keep all cryptocurrency transactions \cite{RN68, RN33, RN69, RN126, RN177}. Blockchain is a public electronic ledger equivalent to a distributed database. It can be openly shared among the disparate users to create an immutable record of their transactions \cite{RN126, RN98, RN81, RN82, RN96, RN36}. Since all the committed records and transactions are immutable in the public ledger, the data are transparent and securely stored in the blockchian network. A blockchain network deploys and executes the programming scripts to process a task autonomously. These programs are called \textbf{smart contracts} which are used to define the customized functions and rules invoked during the transactions \cite{RN110, RN145, RN39}.

Smart contracts based blockchain technology is being embedded into a wide variety of industry applications, such as finance \cite{RN126, RN152, RN53, RN164}, supply chain management, \cite{RN154, RN155, RN156}, health care \cite{RN157, RN158, RN159, RN160}, energy \cite{RN77, RN161, RN162, RN163}, IoT \cite{RN125, RN165, RN166, RN167} and government services \cite{RN126, RN168, RN169}. The financial technology industry has drastically increased the use of blockchain technology and smart contracts executions. It helps reduce infrastructure costs, increase transparency, reduce financial fraud, and improve the time of execution and settlement \cite{RN33, RN39, RN31, RN170}. Some governments in the developing nations are assessing blockchain as a potential replacement for the national currency \cite{RN96, RN97, RN171}. Because of the transparency and traceability features in the blockchain technology, the government can use a permissioned blockchain platform to regulate and analyze how money is flowing in the national financial system \cite{RN82, RN131, RN70}. In the retail and manufacturing industries, blockchain technology helps deliver a better supply chain management and payment with digital currencies in a secure manner \cite{RN172, RN173}. Blockchain allows patients to access their healthcare records securely without a third party verifier \cite{RN157, RN174}. By digitizing the maritime network, the shipping industry can use a blockchain-based ledger to track millions of shipping containers in the ocean \cite{RN175, RN176, RN178}. 

\begin{figure*}[!ht]
\centering
\fbox{\includegraphics[width=0.98\textwidth]{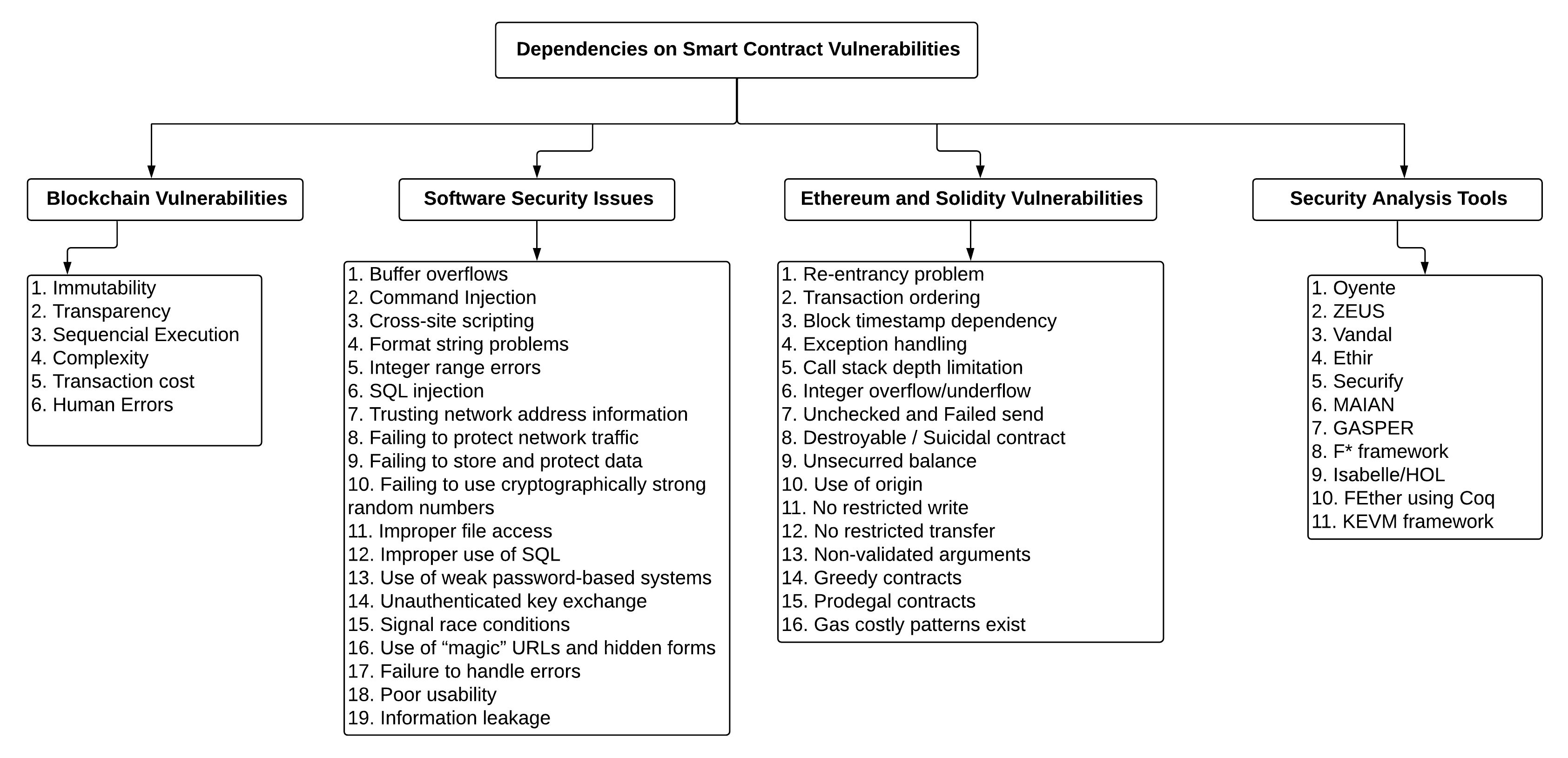}}
\caption{The taxonomy of dependencies in smart contract vulnerabilities}
\label{fig:smartcontract_taxonomy}
\end{figure*}

There are only specific blockchain platforms support smart contracts: Ethereum \cite{RN89} was the first to support smart contracts; other blockchain platforms, such as EOS \cite{RN92}, Lisk \cite{RN91}, Bitcoin \cite{RN90}, RootStock \cite{RN93}, and Hyperledger Fabric \cite{RN94}, are compatible to deploy and execute the smart contracts. A script type language called \textbf{Solidity} is used to develop smart contracts in Ethereum platform. This paper focuses on smart contracts on the Ethereum network. Smart contracts facilitate to develop decentralized applications and perform credible transactions without third parties. Following the pre-defined rules, smart contracts provide trustworthy services as an intermediary during the execution of the transactions. All smart contracts are stored in a distributed consensus environment. That is, once they are deployed to the network, nobody can modify them so that the functions in the deployed smart contracts are immutable. In Ethereum, smart contracts are considered as an account --- they can hold cryptocurrency and transfer between externally owned user accounts and other smart contracts \cite{RN61, RN64}. Since the deployed smart contracts often hold a significant amount of coins \cite{RN98} and perform critical functions \cite{RN145}, they should be tested and analysed before the deployment \cite{RN76, RN108}. However, there are several challenges in smart contracts development using Solidity language: 
\begin{itemize}
\item Users and developers have lack of knowledge about the usage and implementation of smart contracts since the technology is still in an early stage \cite{RN179}. 
\item There are limitation of defined best practices for the programming and testing methods \cite{RN180}. 
\item If any errors identified or detected after the deployment of smart contracts, they cannot be patched and redeployed in the same manner of traditional software updates \cite{RN169, RN142, RN182}. On the contrary, the erroneous smart contract are usually terminated by the owner before an updated contract is deployed.
\end{itemize}

Considering these challenges and issues in smart contract programs in Ethereum, we have come with the following key research questions. 
\begin{itemize}
    \item What are the major attacks occurred in Ethereum smart contracts applications, which caused significant worth of loss in crypto assets?
    \item How are the vulnerabilities in smart contracts affect the systems and how are they exploited by the attackers during the attacks?
    \item What are the security analysis methods available to validate and verify the problems in smart contract programs?
\end{itemize}

There are several security analysis tools and formal verification methods for identifying the vulnerabilities in smart contracts in Ethereum \cite{RN49, RN86, RN28, RN121, RN116, RN118, RN119, RN129, RN80, RN47, RN115, RN129, RN130, RN140, RN220}. They used different types of technical methods to implement their security analysis on smart contracts bytecode or source code. The existing surveys were often conducted in general with the comparison of a limited number of software tools with their coverage of important vulnerabilities \cite{RN50, RN73, RN74, RN111, RN120, RN117, RN214}. Only very few surveys investigate the challenges and security problems in the whole blockchain system \cite{RN124, RN74}. The three most important surveys are listed below: 
\begin{itemize}
 \item Atzei et al.~\cite{RN50} surveyed the past security attacks and possible challenges on Ethereum smart contracts.  
 \item An empirical analysis of smart contracts was conducted by Bartoletti et al.~\cite{RN73}. 
 \item Li et al.~\cite{RN74} reviewed the security of blockchain systems. The security issues of smart contracts in Ethereum network were analyzed in the risk perspectives.  
\end{itemize}

Different from the existing surveys \cite{RN50, RN73, RN74, RN111, RN120, RN124, RN206, RN213, RN243, RN72}, this paper aims to specifically analyse the vulnerability detection methods for Ethereum smart contracts in the context of the identified security attacks \cite{RN65, RN66, RN50}. The taxonomy of dependencies in smart contract vulnerabilities are illustrated in Figure \ref{fig:smartcontract_taxonomy}. We identify the needs of a comprehensive study on security analysis methods \cite{RN49, RN140} of vulnerable smart contracts on Ethereum platform. Moreover, this paper is different from the existing security surveys because we investigate the specific security problems of smart contracts. The major contributions of this survey are as follows:
\begin{itemize}
    \item We identify the security problems and vulnerabilities in Ethereum smart contracts which have caused severe attacks \cite{RN65, RN66, RN50}, and significant loss of cryptocurrency \cite{RN183}. 
    \item We categorize the existing security analysis methods in terms of static analysis \cite{RN49, RN28, RN121, RN116, RN119, RN47, RN120}, dynamic analysis \cite{RN86, RN80}, and formal verification methods \cite{RN45, RN118, RN113, RN148, RN115, RN128, RN129, RN130, RN140, RN149, RN117, RN221, RN240}. 
    \item We compare the analysis methods with the vulnerability findings \cite{RN142, RN49, RN50, RN111, RN124, RN242}, and coverage using their applications, such as vulnerability scanning tools \cite{RN49, RN28, RN121, RN116, RN119, RN47, RN86} and verification models \cite{RN45, RN118, RN113, RN148, RN117}.
\end{itemize}

This survey selects and presents the papers published in high quality journals and presented at the top conference. The keywords we used to search are ``blockchain", ``Ethereum", ``smart contract", ``security analysis", and ``vulnerability". Around 125 research papers from best quality journals, transactions and conferences are included in our survey.

The rest of this survey is organized as follows: Section \ref{sec:II} introduces the basic theory of the Ethereum network and smart contracts. Section \ref{sec:III} covers the major attacks occurred on Ethereum smart contract applications in the recent years. Section \ref{sec:IV} lists the important vulnerabilities in Ethereum smart contracts with respect to the related attacks. Section \ref{sec:V}  presents different types of security analysis methods of smart contracts. Section \ref{sec:VI} compares the analysis methods using their applications and providing a summary of vulnerability identification and possible solutions. Section \ref{sec:VII} covers the research challenges, future research direction and conclusions of this survey.

\section{Background Information}
\label{sec:II}

This section briefly provides the theoretical knowledge of Ethereum platform, Ethereum accounts, and the execution of the Ethereum smart contracts.

\subsection{Ethereum Platform}
Ethereum \cite{RN98} is an open software platform based on the blockchain technology. The developers can implement, compile, test, deploy, and execute the centralized applications in Ethereum network. The Ethereum Virtual Machine (EVM) \cite{RN61} is an abstract machine designed to serve as a run-time environment for Ethereum smart contracts. EVM runs as an independent process on a server or a computer. An Ethereum network is a distributed and decentralized network with permission-less untrusted peers \cite{RN184, RN185, RN186}.

Ethereum network has two types of accounts: One is externally owned user account controlled by the private key, and the other one is smart contract account controlled by its compiled programming code \cite{RN187}. User accounts contain no code and can send messages to other accounts by creating and signing a transaction using their private keys \cite{RN61}. The recipient account can identify the sender by using the sender's public key. Like autonomous agents, contract accounts in Ethereum always execute a specific sequence of code according to the pre-defined rules when the smart contracts are invoked by a transaction \cite{RN108}. 

Each Ethereum account is 20 bytes long \cite{RN61}, and it consists of a unique address, the current balance in Ether, the data storage, and a nonce \cite{RN187}. A nonce is a counter ensuring that each transaction can only be executed once. Ether is the primary cryptocurrency denomination in Ethereum which is used to process the transactions and pay the transaction fees.

\subsection{Smart Contracts}
Smart contracts in Ethereum are computer programs written by programming language called `Solidity' \cite{RN179, RN87, RN188}. The compiled bytecode are deployed in EVM. Any rules and functionalities can be written using compatible programming language and encoded as a smart contract to invoke whenever an action is required by users or other smart contracts. They can implement various kinds of applications of financial instruments such as cryptocurrency management (\emph{ABCC}, \emph{AlterDice} \cite{RN195}), crypto wallets (e.g., \emph{MyEtherWallet} \cite{RN192}, \emph{MetaMask} \cite{RN193}, and \emph{MyCrypto} \cite{RN194}), and autonomous governance applications \cite{RN182, RN190}. Smart contracts are called by users by referring transactions to the contract address. If the transaction is agreed across the network, all the peers have to execute the contract code with the current state of the blockchain with the relevant input parameters \cite{RN182}. 

\begin{figure}[!ht]
\begin{center}
\includegraphics[width=0.47\textwidth]{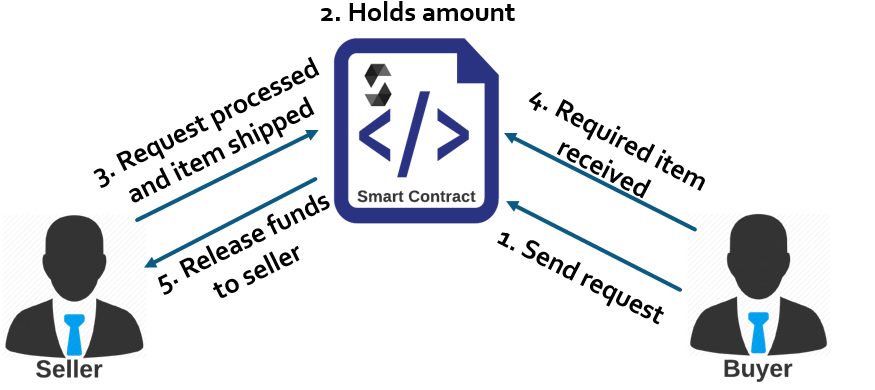}
\caption{A Real-world Example for Smart Contract Execution}
\label{fig:smteg}
\end{center}
\end{figure}

The \emph{Ethereum Network} \cite{RN98, RN89}, one of the leading blockchain platforms, supports the execution of smart contracts enforced by the consensus protocol. \emph{Etherscan} \cite{RN62} is an analytic platform of Ethereum which used to explore blocks, accounts, transactions and statistic data. More than 1,000,000 smart contracts have been deployed in Ethereum platform. We consider a real-world example where we can see how smart contract acts as trusted intermediary between the users. As shown in Figure \ref{fig:smteg}, two users (a seller and a buyer) do business through an application with a smart contract. The whole transaction completes in five steps: In step 1, the buyer sends the required amount of Ethers to the smart contract's address so that the smart contract holds the balance in escrow. In step 2, the smart contract notifies to the seller by triggering an event indicating the recipient of the buyer's request. In step 3, the seller checks and verifies the buyer's request --- if the request is valid and there are enough Ethers to purchase the required item, then the seller will ship the item and inform the smart contract with the shipment message. In step 4, after the buyer receives the item, the smart contract is updated with the delivery status. In step 5, the smart contract releases the Ethers to the seller's account. 

\subsection{Software Security Vulnerabilities}
Howard et al.~\cite{RN228} classified the 19 software security issues and how they affect the software programs in different ways. These are the major software vulnerability categories and followed and cited by plenty of researchers \cite{RN229}. We referred these 19 vulnerability categories and mapped them with Ethereum security vulnerabilities in our analysis. 

\section{Unique Security Attacks against Smart Contracts}
\label{sec:III}

This section covers the important security attacks and relevant vulnerabilities of smart contracts implementation on the Ethereum platform. Smart contracts can hold and manage a large amount of virtual currencies which could be worth thousands of dollars \cite{RN49,RN169}. Therefore, the adversaries keep attempting to manipulate the execution of smart contracts in favor of their activities. In nature, smart contracts are running on the distributed and permission-less networks, which inherits many vulnerabilities \cite{RN49, RN50}. The attacks occurred due to the malfunctioning of the smart contract execution lead to a massive amount of loss in virtual currencies \cite{RN65, RN66}. In a traditional system, the buggy applications running in a centralized environment can be redeveloped or patched \cite{RN196}. On the contrary, in a decentralized blockchain network, the deployed smart contracts cannot be modified or upgraded in a live network unless extreme measures are taken \cite{RN169, RN109}. The immutable nature of smart contracts makes pros and cons in the means of security aspects. Because of this immutablity, hackers are unable to make changes or modify the contracts for their benefit. However, the smart contract applications cannot be modified even by the developers after the deployment. They can kill or terminate the contract and create new smart contract and deploy it again. Therefore, before the deployment, the smart contracts should be thoroughly tested with a wide range of test cases for security and safety reasons.  

\subsection{The DAO Attack}

In June 2016, the DAO hack occurred when the attacker managed to steal more than 3.6 million Ethers \cite{RN65}. The DAO attack was caused by a re-entrancy problem \cite{RN49, RN50} existing in the smart contract. The re-entrancy problem allowed the attacker to exhaustively execute recursive calls for requesting and receiving funds from the vulnerable \texttt{DAO.sol} contract listed below. That is, the attacker kept withdrawing Ethers by requesting the DAO smart contract before updating the balance of smart contract. The \texttt{withdraw} function of the target contract (\texttt{DAO.sol}) was called recursively until the contract balance reached zero. More specifically, the attacker embedded the \texttt{withdraw} function in a fallback function of the smart contract \texttt{DAOAttacker.sol}. The fallback function is a default function in the Ethereum smart contracts and can be declared without any explicit function name. Because the fallback function is automatically called whenever the attacker receives any funds, the smart contract inherently calls the embedded \texttt{withdraw} function. This setup allowed the attackers to call the \texttt{withdraw} function recursively before the user's balance is updated, e.g., before sending any funds.

\begin{table*}[!ht]
  \begin{center}
  \caption{Security Attacks on Smart Contracts and Available Solutions}
\label{tab:security_attacks}
\begin{tabular}{ |p{0.2\textwidth}|p{0.22\textwidth}|p{0.3\textwidth}|p{0.18\textwidth}|} 
\hline
Major Attacks & Ethereum Vulnerabilities & Available Solutions & Software Security Issues \\
\hline
\multirow{4}{12em}{The DAO attack (2016) \cite{RN65}} & Re-entrancy on a single function  &  Use \texttt{send()} instead of \texttt{call.value()} & Failing to store and protect data \\
& Re-entrancy on cross functions  &   Use \texttt{send()} or \texttt{transfer()} to send funds  & Race conditions \\
& Re-entrancy on external contract functions &  Do internal state changes first and then call external function; use a mutex when the external calls are unavoidable  & Improper file access \\
\hline

\multirow{3}{12em}{Parity Multi-Sig Wallet Attack (2017) \cite{RN66}} & Public functions are callable by anyone & Use the \texttt{internal} modifier for functions instead of \texttt{public} & Information leakage\\
& (No access modifier assigned properly) & Explicitly define library functions for the external invocations & Improper file access \\
\hline

\multirow{2}{12em}{Over/Under flow attack (2018) \cite{RN50}} & \multirow{2}{*}{Integer underflow and overflow} & Check if the integer stays in its byte range before any send operations & Integer range error, Buffer overflow \\
\hline
\end{tabular}
\end{center}
\end{table*}

\begin{lstlisting}
// DAO.sol
contract DAO {
    // assign Ethers to an address 
    mapping(address => uint256) public deposit;
    
    // credit an amount to sender's account
    function credit(address to) payable {
        deposit[msg.sender] += msg.value;
    }
    
    // get credited amount
    function getCreditedAmount(address) returns (uint) {
        return deposit[msg.sender];
    }
    
    // withdraw fund from contract
    function withdraw(uint amount) {
        if (deposit[msg.sender] >= amount) {
        msg.sender.call.value(amount)();
        deposit[msg.sender] -= amount; }
    }
}
\end{lstlisting}

A sample target contract named \texttt{DAO.sol} and an attacker contract named \texttt{DAOAttacker.sol} are listed to explain the technical details. First, the attacker sends some Ethers to the DAO contract by invoking the credit function in line 7--9 of \texttt{DAO.sol}.  The attacker's balance is updated by the DAO contract according to the amount of Ether in line 8 of \texttt{DAO.sol}. Then, the attacker sends a request to withdraw the fund. According to this withdraw request, the fund is sent back to the attacker's contract (Line 17--21 of \texttt{DAO.sol}). After the funds are received, the fallback function is called for a continuous withdrawal as per line 15 of the contract in \texttt{DAOAttacker.sol}. Since the target contract has not updated the attacker's balance yet, the withdrawal request will be successfully executed. This repeating process ended up stealing the all available funds from the target contract. Finally, the attacker transfers the stolen funds from the \texttt{DAOAttacker.sol} contract to a pre-defined personal account address (Line 20 in the \texttt{DAOAttacker.sol} contract).

\begin{lstlisting}
// DAOAttacker.sol
import 'DAO.sol';
contract DAOAttacker {
    
    // initialize DAO contract instance
    DAO public dao = DAO(0xDa32C9e....);
    address owner;
 
    //set contract creator as owner
    constructor(DaoAttacker) public {
        owner = msg.sender;
    }
    //fallback function calls widthraw function
    function() public { 
        dao.withdraw(dao.getCreditedAmount(this));
    }
    
    /*send stolen funds to attacker's address*/
    function stealFunds() payable public{
        owner.transfer(address(this).balance);
    }
}
\end{lstlisting}

As listed in Table \ref{tab:security_attacks}, the DAO attack caused by a Re-entrancy problem as an Ethereum vulnerability is related to a few software security issues including the improper file access problem, race condition issue, and failing to store and protect data. The solidity programming practice, namely the \texttt{call} method, has caused the attacker to invoke the \texttt{withdraw} method of the \texttt{fallback} function. Since the balance is updated after invoking the \texttt{call} method, the data is not properly stored or protected at the correct time. The intermediate state of the data or balance was taken and mistreated by the attacker to his beneficiary action. The actual problem is entirely caused by a smart contract programming error not by the Ethereum network. Any network which had this type of erroneous smart contact would facilitate the re-entrance hack.

For the immediate solution for this attack, there were many arguments of deciding how to refund the funds to the victim and terminate the hacked DAO contract.  The hard fork mechanism overwrites the history of transactions by reversing them to the starting state. However, the hard fork did not prevent all Ethereum users to go along with the old main branch. The Ethereum branch created with hard fork is running as original Ethereum, and the old branch is keep working as the Ethereum Classic \cite{RN230, RN231}. The DAO attack has triggered the Ethereum developers to enforce proper coding regulations and practices on smart contract development because the blockchain's immutability and smart contract's deterministic features are hard to resolve sudden attacks.

\subsection{Parity Multi-Sig Wallet Attack}
The parity multi-sig wallets are smart contract programs which are used to manage digital assets by the wallet users \cite{RN66}. The important data, such as daily withdrawal limits, ownership information, and withdrawal voting, are configured and stored in these wallets \cite{RN197}. If a user wants to own a multi-sig wallet, the user should have multiple signatures (that is, private keys) to withdraw funds from the wallet. This signature requirement strengthens the security of the wallet, especially those that are involved in the transactions with significant worth of crypto assets \cite{RN198}. Some of the frequently used functions and logic of the parity multi-sig wallet are implemented in a public library \cite{RN207}. This shared wallet library is available to every parity multi-sig and supports the essential methods, such as withdrawing fund, setting withdrawal limit, depositing fund, and so on. Multi-sig wallets are able to call these external public functions from their contracts \cite{RN198}. The centralized setup of this library becomes a target of attacks. The parity multi-sig wallet attack occurred when the attacker managed to initialize the public library as a multi-sig wallet and subsequently gained the ownership right and the killing right \cite{RN197}. Since all wallets depend on this public library, their deployed contracts were useless against the attacker. Around 151 wallets were frozen with their balances reaching 15,153,037 Ethers in total \cite{RN66}. This attack is the second largest attack on the Ethereum network in terms of the amount of stolen Ethers \cite{RN142}.

\begin{lstlisting}
// WalletLibrary.sol
// constructor in wallet library
// set daylimit and muliple owners
function initWallet(address[] owners, uint required, uint dayLimit) {    
  initDaylimit(dayLimit);    
  initMultiowned(owners, required);  
}
\end{lstlisting}

The code snippets of the two contracts are shown as \texttt{WalletLibrary.sol} and \texttt{MulitisigWallet.sol}. The attacker's first transaction was sent to the wallet contract to claim the ownership of the multi-sig wallet. The second transaction was sent to withdraw all the funds from the wallet. In the contract  \texttt{WalletLibrary.sol}, the \texttt{initWallet} function initializes a wallet with the parameters of day limit, array of owners or signers and the required number that needed to confirm a transaction. This is a constructor written in the external wallet library and it is publicly available for invocation by anyone using the delegate calls \cite{RN142}. After the attacker claims the ownership with the multi-sig wallet, all the funds available in the wallet can be stolen \cite{RN66}. The function \texttt{delegatecall} is called by a wallet instance as in Line 8 of \texttt{WalletContract.sol}. The main problem caused by this attack is that all the public functions, such as \texttt{initDayLimit} and \texttt{initMulitowned}, in the \texttt{WalletLibrary.sol} contract can be called by anyone without authorization. There was no access modifier used to restrict the invocations from anonymous callers. The modifiers \texttt{internal} or \texttt{private} can be used for the functions to be called within a contract or from derived contracts \cite{RN199}.

\begin{lstlisting}
// MultisigWallet.sol
function() payable {
  // deposit an amount to sender's address
  // walletLibrary is an instance of the public library
  if (msg.value > 0)
    Deposit(msg.sender, msg.value);
  else if (msg.data.length > 0)
    walletLibrary.delegatecall(msg.data);
}
\end{lstlisting}

The parity-multisig wallet attack was related to a few software security issues including improper file access and information leakage problem according, as shown in Table \ref{tab:security_attacks}. The call to an external library caused the problem since the library function did not have proper access control. Thus, the attack mainly focused on the weak library and non-restricted invocations to the external wallet library functions. The non-updatable nature of the blockchain enables the attackers to target the problematic libraries as well as smart contracts to attack the smart contract applications. The initialization logics were developed in the library constructor. Despite this concept of abstraction is good for re-usablitiy, it facilitates the hackers to invoke a call \texttt{delegatecall} to the library functions and gain the full control of the library. 

The majority of parity users did not agree to perform another hard folk for refunding the locked Ethers from the affected wallets \cite{RN232}. The hard folk applied in the DAO attack split the Ethereum network into two networks, and the hackers' stolen funds are still valid in the Ethereum Classic version \cite{RN232}. A white-hat recovery team promised to provide a new parity wallet for each affected wallet with the restored settings same as the ones before the attack. They could recover the remaining fund in the frozen wallets and remove the vulnerability from the wallet contracts. Afterwards, it is recommended for Solidty developers to adopt the \texttt{private} modifier by default to restrict the access for all contract functions \cite{RN233}. This restriction will disable the malicious function calls to wallet library functions by anonymous users.

\subsection{Integer Overflow/Underflow Attack}

The Proof-of-Week-Hands (POWH) Coin is a Ponzi scheme developed by a group of people using smart contracts. It had been attacked due to an integer overflow/underflow problem in 2018. The attacker drained around 2,000 Ethers because of the insecure operations of integers \cite{RN200}. An unsigned integer in Solidity is defined as \texttt{uint256} \cite{RN87}. Each \texttt{uint256} is limited to 256 bits in size translating to any integers between 0 and 4,294,967,295 ($2^{256}-1$). If an integer variable assigned to a value larger than this range, it resets to 0; if the variable assigned to a value less than the range, it would be reset to the top value of the range \cite{RN87}. For example, when a positive number is subtracted from 0 it will result an integer of $2^{256}-1$. The attacker exploited this vulnerability to steal Ethers through such an integer underflow attack \cite{RN50}. 

If an attacker has a target account holding 0 Ether, an attack example works as the following steps:  First, the attacker sends 1 Wei to a target contract. (Wei is the smallest denomination of Ether in Ethereum --- 1 Ether is worth $10^{18}$ Weis \cite{RN201}.) The target contract will deposit the fund to the sender's account. Next, the attacker requests to withdraw 1 Wei, and the sender's balance will be updated to 0 Wei by subtracting 1 Wei. When the target contract sends the fund to attacker's contract, the attacker's fallback function will be triggered so that a subsequent withdrawal is requested again. Now when the contract updates the balance by subtracting 1 from 0, the balance becomes -1. Due to the integer under/over flow issue, the attacker's balance will be automatically reset to 2 Weis.  Using a repeating mechanism similar to the re-entrancy problem in the DAO contract, the attacker is able to steal all funds from the victim's account.

Furthermore, the solidity compiler does not trigger any error flag to resolve the code with integer overflow/underflow problems. The integer overflow/underflow problem can be mitigated through using the arithmetic functions in the Solidity math library named \texttt{SafeMath.sol} \cite{RN234}. It supports safe mathematics operations, such as addition, subtraction, and multiplication, while preventing the integer overflow/underflow issues. 

Solidity language is less flexible since it has limitations on the value/integer types and length \cite{RN69}. Several memory error detection techniques have been proposed for C and C++: The \emph{StackGuard} automatic buffer overflow detection \cite{RN235}, \emph{PointGuard} protection \cite{RN236}, baggy bounds checking \cite{RN237}, and the light weight bounds checking \cite{RN238} are popular choices for bounds checking  C and C++ programs. Since these bounds checking problems exist widely in Solidity language, prevention mechanisms should be developed to perform proper bounds checking as in C and C++. An overflow detector named \emph{EasyFlow} \cite{RN239} can identify the manifested overflows, well-protected overflows and potential overflows in vulnerable smart contracts.

\subsection{The Learned Lessons}

According to our analysis on the major attacks occurred on Ethereum smart contracts, the Parity multisig wallet attack made severe impacts to the Ethereum by freezing a massive amount of funds, even though the attack was technically simple. The vulnerability was affected in both wallet contract and external library contract. It is challenging to detect the deployed libraries that leak the information and set inappropriate level of the control without proper access modifiers. These library contracts can self-destruct caused by malicious users with an escalated privilege. These attacks are simple and straight-forward because it is obviously abnormal to lock or freeze the smart contracts holding a significant amount of funds after a function call. The erroneous or vulnerable contracts are deployed to the Ethereum network without proper security checks, quality assurance tests, or following the best coding practices in Solidity. 

The combination of vulnerabilities in Ethereum blockchain and Solidity programming language makes the security checks more challenging in smart contracts development \cite{RN179}. Compared to native languages like Java, C and C++, the Solidity language is not very mature as a scripting language. Since integer types are fixed in size with 256 bits, the buffer overflow/underflow bugs in Solidity make erroneous smart contracts. Furthermore, the \texttt{mapping} data type in Solidity will not throw exception even if there is no key-value pair, instead it simply returns the default value. This nature can allow the attackers to execute the malicious codes by passing the parameters to the attackers' advantage into smart contract functions with the \texttt{mapping} data type. Since Solidity functions can be recursively called, it lacks the tail call support \cite{RN136}. Thus, the depth of recursive calls can be defined exclusively through input variables of the smart contracts.

In addition to the well-known attacks, there are more vulnerabilities in smart contracts. Many of them are proven to be problematic. They make less impact than the attacks, but they present a landscape of the security issues of smart contracts which is investigated in Section \ref{sec:IV}.

\section{Key Vulnerabilities in Smart Contracts}
\label{sec:IV}

In this section, we discuss the key vulnerabilities which would cause serious problems in smart contracts applications. Re-entrancy problem, Transaction ordering dependency problem, Timestamp dependency problem and Exception handling issues are causing vulnerable patterns in smart contract execution as well as in their code. Developers should aware of these issues and have to follow quality assurance test cases carefully before they deploy their contracts into live Ethereum or any blockchain platform. Further we investigated 16 Ethereum vulnerabilities as shown in Table \ref{tab:vulnerabilities_attacks_sins}. It describes Ethereum vulnerabilities and their related attacks. Also it maps relevant software security issues as categorized in \cite{RN228} with the identified key Ethereum vulnerabilities.

Since smart contracts are executing asynchronously, the transaction ordering problem is a common attack vector. This problem can be cured using a locking mechanism which will keep an order or counter for each transaction to execute by first-in-first-out manner. Timestamp dependence problem is a prominent issue that uses block timestamp in critical operations. It is recommended to avoid assigning block timestamp to a variable in smart contract code. Instead of timestamp value, block number can be used for a constant variable. Exception handling problem is one of major problem in solidity programming. Developers can handle this problem by having best practices and exception try-catch mechanisms. The latest versions of solidity compiler also aware of this issue and giving warning or error message when compiling a code without having a proper exception handling implementation. 

\subsection{Re-entrancy Problem}
As illustrated in Section \ref{sec:III}.A, the DAO attack was occurred due to re-entrancy problem in smart contracts. The solidity smart contract has an unnamed function called \emph{fallback} function that does not have any arguments nor return values. The \emph{call} function is used to invoke a method of external contract or the same contract to transfer Ethers. This function does not throw any exception if any errors prompted, but it returns false otherwise true. This \emph{call} method executes without a gas limit if it has not being set any gas value manually. If a contract invoke a \emph{call} method to send an amount to sender's account, it will call sender's \emph{fallback} function. Since there is no gas limitation for \emph{call} method invocation, any code inside the \emph{fallback} function would be executed until it finishes the remaining gas amount. This vulnerability is called re-entrancy in Ethereum smart contract and it was the serious attack vector for the DAO attack. A dynamic analyzing tool called \emph{ReGuard} \cite{RN241} detects the re-entrancy problem in smart contracts with the identification of unknown problems.

\begin{table*}[!ht]
  \centering
  \caption{Relating the Ethereum Vulnerabilities, Major Attacks, and Relevant Software Security Issues}
 \label{tab:vulnerabilities_attacks_sins}
\begin{tabular}{|p{0.18\textwidth}|p{0.3\textwidth}|p{0.2\textwidth}|p{0.2\textwidth}|}
\hline
\textbf{Ethereum Vulnerabilities} & \textbf{Vulnerability Mechanism} & \textbf{Related Attacks} & \textbf{Software Security Issues}\\
\hline
Re-entrancy problem & Recursively calling a function from a \texttt{fallback} function & The DAO attack &  Failing to store and protect data \\
\hline
Transaction ordering & Inconsistent transactions' orders  with respect to the time of invocations & - & Race conditions\\
\hline
Block timestamp dependency & Constant variables are assigned to block timestamp value & - & failing to use cryptographically strong random numbers \\
\hline
Exception handling & Failing to check the return values  after a function call &
The DAO attack, Integer Over/Under flow attack, King of Ether Throne attack & Failure to handle errors \\
\hline
Call stack depth limitation & Exceeding the limit of number of calls to a contract method & - & Buffer overflows \\
\hline
Integer overflow/underflow & Subtracting positive integers from zero results big value  & Integer Over/Under flow attack & Integer range errors\\
\hline
Unchecked and failed \texttt{send} & Send Ethers without checking the conditions  & The DAO attack & Failing to store and protect data, Failure to handle errors \\
\hline
Destroyable / suicidal contract & Contract is susceptible to be destroyed by unauthorized users & Parity Multisig Wallet attack & Improper file access \\
\hline
Unsecured balance & The Ether balance in a contract is exposed because of the modifier \texttt{public} to theft by an anonymous caller & The DAO attack, Parity Multisig Wallet attack & Failing to store and protect data \\
\hline
Misuse of \texttt{ORIGIN} & Contract authenticates using the return value of \texttt{ORIGIN} rather than \texttt{CALLER} & - & Failing to store and protect data \\
\hline
No restricted write &  Writes to storage variable is restricted by the modifier \texttt{private}  & Parity Multisig wallet attack & Failure to store and protect data \\
\hline
No restricted transfer &  Ether transfers cannot be invoked by any user who is independent to the sender  & The DAO attack, Parity Multisig wallet attack & Failure to store and protect data \\
\hline
Non-validated arguments & Arguments in a contract function should be validated before its use  & Integer Over/Under flow attack & Failure to handle errors \\
\hline
Greedy contract & Locking the contract fund or Ether balance indefinitely  & Parity Multisig Wallet attack & Improper file access, Failure to store and protect data \\
\hline
Prodigal contract & Leaking fund or Ether balance to arbitrary users  & The DAO attack & Information leakage\\
\hline
Gas overspent & Contract code execution consumes more gas unnecessarily & - & Poor usability \\
\hline
\end{tabular}
\end{table*}

\subsection{Transaction Ordering Dependency}
A block includes a set of transactions, and the blockchain state is updated several times during each epoch \cite{RN49, RN74}. The state of a smart contract is jointly determined by the value of its fields and the current balance \cite{RN202}. In most cases, when a user initiates a transaction to invoke a smart contract in the network, there is no guarantee on whether the transaction will run in the same state that the contract was at the time of the initialization of the  transaction. The actual state of the smart contract is unpredictable by any user when it was called by the user's transaction \cite{RN49, RN50, RN74}.

\begin{figure}[!ht]
\centering
\fbox{\includegraphics[width=0.47\textwidth]{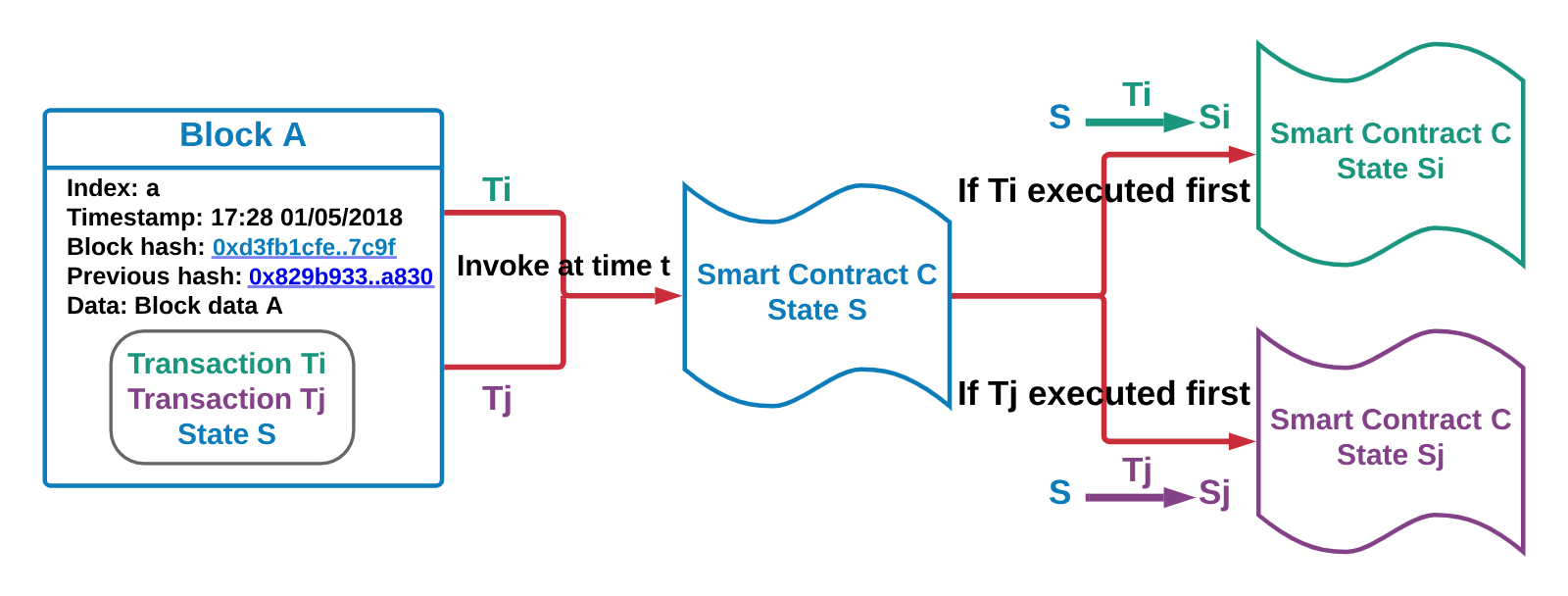} }
\caption{The Transaction Ordering Dependency Problem \cite{RN49,RN50}}
\label{fig:order}
\end{figure}

If a new block on a blockchain includes two transactions to invoke the same contract, then the users have no certain knowledge of which state the contract is at when their individual invocation is executed. As shown in Figure \ref{fig:order}, if user1 and user2 respectively send transaction $T_{i}$ and $T_{j}$ to a smart contract at same time $t$, both users do not know which transaction will first run. And the order of these transactions are determined only by the miners of the block. Even if user1 sends transaction $T_{i}$ before user2 sends $T_{j}$, $T_{i}$ is not guaranteed to run before $T_{j}$. If $T_{i}$ is executed first, it will change the contract state from state $S$ to state $S_{i}$; but if the $T_{j}$ is executed first, it will change the contract state from state $S$ to state $S_{j}$. Therefore, the final state of a contract depends on the order of transaction execution which is determined by the block mining order. 

This problem is critical in the real-world situations where buyers and sellers use smart contracts for their decentralized stock market operations as implemented in the \texttt{StockMarket.sol} contract shown below. Sellers will often update the price of their selling items, and buyers will send their orders to purchase those items with the expectations of the price as they observed when they sent the transaction. In the worst case scenario, buyers may have to spend significantly more than their expected price for the requested item.

\begin{lstlisting}
// StockMarket.sol 
contract StockMarket {
    uint public stock_price;
    uint public stock_available;
    address public owner;
    
    function updatePrice (uint _price) private {
        if(msg.sender == owner){
            stock_price = _price
        }
    }
    
    function buy (uint quantity) private returns (uint) {
        if(msg.value < quantity*stock_price || quantity > stock_available)
        stock_available -= quantity;
    }
}
\end{lstlisting}

\subsection{Timestamp Dependency}

The smart contract uses the block timestamp as an initial condition to execute some critical operations. Usually the timestamp is set to the system time of the miner's local computer or server \cite{RN49,RN50}. When a block is mined, the miner has to generate the timestamp for the block. The timestamp of a block can vary by approximately 900 seconds comparing with other blocks' timestamps \cite{RN49, RN201}. If a miner received a new block after the validity conditions are confirmed, the miner will check whether the timestamp of the received block is greater than the timestamp of previous block and whether his local machine timestamp is not greater than 900 seconds from the received block's timestamp \cite{RN49}. Because of this flexibility in setting the timestamp of a block by miners, an adversary or malicious miner can choose different block timestamps to manipulate the outcome of timestamp dependent smart contracts. If a contract is using the current time ($now$), starting time ($StartTime$) and ending time ($EndTime$) based on the timestamp of the block, that means that the miner can manipulate the timestamp for a few seconds by changing the output for the miner's favor \cite{RN49, RN50}.

The following code snippet of \texttt{TheRun.sol} contract uses the block's timestamp value to generate a random number which is subsequently used in a critical operation for the calculation. In line 2, a private variable $salt$ is assigned to the timestamp of the block as a random number. In the \texttt{random} function, the $salt$ variable is used to calculate the values of parameters $x$, $y$, and $seed$. And it returns the calculated number whenever the function is externally called. 

\begin{lstlisting}
// TheRun.sol -- function random()
uint256 constant private salt = block.timestamp;

function random(uint Max) constant private returns (uint256 result){
    //get the best seed for randomness
    uint256 x = salt * 100 /Max;
    uint256 y = salt * block.number / (salt %5) ;
    uint256 seed = block.number/3 + (salt % 300) + Last_Payout + y;
    uint256 h = uint256(block.blockhash(seed));
    
    return uint256((h/x)) % Max + 1 // random number between 1 and Max
}
\end{lstlisting}

The following code implements the condition where the random function is called in line 4. The return value of random function is calculated by the block's timestamp and assigned to the variable $roll$. Then the variable $roll$ is checked for a condition --- if it is successful, then it will run the \texttt{send} function as a critical call. A malicious miner can take advantages by modifying the local system's timestamp to trigger this call. 

\begin{lstlisting}
//TheRun.sol -- call random() function
//winning condition with deposit > 2 and having luck
if( (deposit > 1 ether ) && (deposit > players[Payout_id].payout) ){
    uint roll = random(100); // create a random number
    if( roll % 10 == 0 ) { 
        msg.sender.send(WinningPot);
        WinningPot=0;
    }
}
\end{lstlisting}

Similarly, there are smart contracts which use the block hash value on crucial components. It is not recommended, because the malicious miners can still manipulate the timestamp in order to modify the execution output.

\subsection{Mishandled Exception Issues}
In Ethereum, a smart contract often needs to call another to fulfill the required functionalities \cite{RN108}. These calls are conducted by either sending instructions or calling a contract's method directly with reference to the contract's name \cite{RN49,RN50}. In the callee contract, there may be exceptions raised so that the callee contract will terminate and revert its state while returning a false value to the caller contract \cite{RN49,RN50}. The exceptions can be caused by many situations, such as there is not enough gas to execute the operation, the call stack limit is exceeded, some unexpected system error occurs in the callee node, and so on \cite{RN49, RN203}. The exception thrown in the callee contract should be propagated to the caller, and the return value should be explicitly checked in the caller contract to verify whether the call has been executed successfully or not \cite{RN39, RN49, RN50, RN118}. In several instances of smart contract calls, there are inconsistencies in the exception propagation policies \cite{RN49}, which posts threats in the real-world transaction.

A malicious user can invoke a caller contract and cause its \texttt{send} function to fail purposefully. The call-stack depth is the maximum time a function can be called iteratively \cite{RN108, RN49, RN50}. The Ethereum Virtual Machine sets the call-stack depth limit to 1,024 frames \cite{RN98}. If the 1024-frames limit is exceeded, the EVM will throw an error. The value of the call-stack depth is increased by one if a function is called at once. An attacker can use this feature to intentionally interrupt the execution by calling a contract itself for 1,023 times \cite{RN108, RN98, RN49}.

An example of a contract which is vulnerable to the call-stack depth exceed problem is a Ponzi scheme implementation \cite{RN62}. The \texttt{SimplePonzi.sol} contract is shown in the following code snippet.  This contract is used to pay interest to the investors according to their amount of investments and the order of the investments. An attacker can exploit the call-stack limit to gain benefit by getting his/her interest earlier. And the attacker can intentionally make other investors payments fail by increasing the call stack depth to 1,023. Having executing these calls, the attacker will make his/her payment to receive the interest earlier than other investors since their payments are terminated or unsuccessful.   

\begin{lstlisting}
//SimplePoinzi.sol
contract SimplePonzi {
    address public currentInvestor;
    uint public currentInvestment = 0;
    
    function() payable public {
        unit minimumInvestment = currentInvestment * 11 / 10;
        require(msg.value > minimumInvestment);
        
        //document new investor
        address previousInvestor = currentInvestor;
        currentInvestor = msg.sender;
        currentInvestment = msg.value;
        
        //payout previous investor
        previousInvestor.send(msg.value);
    }
}
\end{lstlisting}

According to the Ethereum documentation \cite{RN201}, using the \texttt{send} function is dangerous and causes many problems. For instance, a transfer fails if the call-stack depth is over 1,024 frames that can be deliberately forced by a malicious caller; and it fails if the recipient runs out of gas. Therefore, in order to safeguard Ether transfers, the return value of any function call should be always checked \cite{RN108}. It can be any invocation of functions used in the contract itself or another contract \cite{RN201, RN49, RN50, RN108}. To prevent the unchecked-send bug \cite{RN50, RN28}, the error should be handled in the caller statement manually; otherwise, it can lead an attacker to execute the unwanted or malicious codes into the contract to rob off its balance.

\subsection{Sequential Execution of Smart Contracts}
Blockchain network such as Ethereum supports the sequential execution of transactions on smart contracts with a consensus mechanism \cite{RN179, RN40, RN29, RN204, RN39, RN179}. In a sequential execution, the requests to the smart contract invocations are ordered by the consensus method. Then, the smart contracts are executed in the same order on all the nodes. This method has many performance limitations and drawbacks in the blockchain-based applications \cite{RN70}. In particular, the most severe problem is that effective throughput of blockchain application is affected due to the sequential operations. The throughput is inversely proportional to the latency of execution \cite{RN82}, which causes the performance bottleneck. Hence, a malicious user can try to introduce a smart contract which may take very long time for its execution. This action will subvert the performance of the network by delaying the traffic of subsequent transactions. 

The sequential execution of smart contracts causes the performance issues by limiting the number of contracts executed per second. The performance in the execution rate of transaction will affect by the sequential execution pattern. The number of smart contracts that can be executed per second will be limited. Vukoli\'{c} et al.~\cite{RN70} proposed to execute the independent smart contracts in parallel to significantly improve the throughput of the transactions. Furthermore, the blockchain-based applications could not be scaled with the growing number of smart contracts in the future \cite{RN205}.

\subsection{Other Ethereum Vulnerabilities}
\textit{Call stack depth limitation:} The call stack depth limit is 1,024 frames in the EVM implementation. When a contract invokes a \texttt{call} or \texttt{send} function to call another contract, the call stack depth increases by one. This setup allows an attacker to exploit a contract by calling itself for 1,023 times before invoking a \texttt{send} function, which exceeds the call stack depth limit \cite{RN49}. The attack exploited on the \texttt{KingOfEtherThrone} smart contract (KoET) due to the call stack depth limit purposefully exceeded by calling the attacker's contract 1,023 times before invoking a \texttt{call} function to claim the throne.

\textit{Integer overflow/underflow:} The integer type \texttt{unit256} in Solidity has a limited size up to 256 bits. If the value of integer variable reaches its maximum value as $2^{256}-1$, then it will automatically be reset to zero when an additional integer 1 is added to the variable. Hackers are keen to target these variables in smart contract to make vulnerable by increase or decrease the value of integers until they reach to the maximum or minimum value \cite{RN244}.

\textit{Unchecked and failed \texttt{send}:} The use of \texttt{send} instruction to send money to another contract or user may fail to send the value to the recipient for reasons like exceeding gas limit or the insufficient amount of Ether in balance. But it will not throw any exception or error message to the contract. If there is no exception handling implemented at invoking \texttt{send} method, the balance would be updated as if it has been sent.

\textit{Destroyable contract:} A destroyable contract \cite{RN34} refers to the smart contract subject to be terminated or killed by an anonymous \texttt{suicide} instruction called by any external user account or another smart contract. The self-destruct function in the smart contract is usually executed by its owner whenever an attack or emergency incident is detected. The self-destruct function should be aware of the user who is executing it, and it should allow the \texttt{kill} method invoked by the legitimate owners only.

\textit{Unsecured balance:} If the balance of any smart contract is exposed to be drained off by a hacker or anonymous caller, the contract is vulnerable with unsecured balance. It can be caused by the improper access control mechanism for balance variable and constructor functions or updating balance after invoking \texttt{call} instruction to send money to another contract or arbitrary user \cite{RN34, RN121}.

\textit{Use of \texttt{ORIGIN}:} In an Ethereum Virtual Machine, the account address initiating the transaction is returned by the keyword \texttt{ORIGIN}; the account/contract address executing the current invocation is returned by the keyword \texttt{CALLER} \cite{RN121}. If a contract has a code that validates the authentication of account/contract that invokes the current message call using \texttt{ORIGIN}, then it is prone to be an erroneous contract.

\textit{No Restricted write:} If there is a possible write operation to the storage without any restricted condition, then it allows the attackers to exploit the contract \cite{RN28}. The parity multisig wallet was hacked because of the absence of restricted write to the storage variable. Therefore, the attacker could set the ownership of wallet library without any condition or proper authorization checks \cite{RN66}. 

\textit{No Restricted transfer:} The \texttt{call} method of Ethereum transfers Ethers between accounts or smart contracts. Despite its convenience, it is not the best practice to have \texttt{call} invoked by arbitrary users. The contract that has no user restriction of sending Ethers through the \texttt{call} function is vulnerable to no restricted transfer. In the DAO attack, the contract sends Ethers to the withdrawer using the \texttt{call} method. This is one of the causes to invoke a fallback function of the attacker's contract and subsequently drain off the money repeatedly using the re-entrance property.

\textit{Non-validated arguments:} Most Solidity functions in Ethereum smart contracts need a few arguments. The arguments in a function are the parameters passed during an invocation of a method or a transaction. The arguments are used in the method for several operations and computations as the required logic. These method arguments should be checked and validated before passing to the method call since the unchecked arguments may cause malicious actions during the execution of the method.

\textit{Greedy contract:} The smart contracts that are remaining active and keep locking Ether balance continuously due to the inability to access the external library contracts to transfer or send fund. These contracts are defined as greedy contracts according to \cite{RN34}. If the library contracts are terminated or destructed by an arbitrary user either intentionally or accidentally, the contracts that call the external library functions are becoming greedy contract \cite{RN34}. The attackers made the Parity Multisig wallets contracts as greedy contracts by claiming the ownership of the wallet library contracts and subsequently destructed them to freeze the money in the wallet contracts \cite{RN66}.

\textit{Prodegal contract:} Ethereum smart contract functions are used to refund the owners after an attack. They transfer Ethers to the addresses who have sent the fund previously or to whom they have provided a solution for a specific problem. These sending process is saved as transactions and contracts are aware of the recipients. In some cases, the contracts are transferring money to arbitrary recipients who have never intervened with these contracts and no data about those addresses. In this scenario, the contracts which send fund to the anonymous users are called Prodegal contract \cite{RN34}, since their sending function can be invoked by any user to send fund to the list of addresses by the sender's choice.  

\textit{Gas costly pattern exists:} The solidity code in Ethereum smart contracts are implemented with expensive patterns which cost more gas during execution of each instructions. There were seven gas costly patterns in contract code identified in \cite{RN47}. These patterns were detected by a tool called \emph{GASPER}. However, the smart contract developers should be aware of their coding practice and optimize the code before they deploy the contracts to the live Ethereum network. It would save contract user's money from spending more gas for the execution of contract methods.

\section{Security Analysis Methods on Ethereum Smart Contracts}
\label{sec:V}

Smart contracts in Ethereum are autonomously intermediate during the execution of transactions. Although they facilitate the blockchain-based applications, there are many security risks and vulnerabilities in the smart contracts. One of the critical challenges in smart contracts is that they are immutable and cannot be upgraded or patched once deployed to the blockchain network. If users' requirement is changed or any errors is found later on their deployment, they cannot be modified like traditional software applications. Furthermore, it is difficult to test smart contracts during their run-times. Because they interact with other smart contracts and invoke many external off chain services repeatedly and continuously. The attackers are very keen to exploit the bugs on smart contracts since these contracts hold significant value of crypto assets. Their effort would be worth to obtain much benefits by stealing fund from smart contracts.

\begin{table}[!ht]
  \centering
  \caption{Types of Security Analysis Methods of Ethereum Smart Contracts}
 \label{tab:analysis_methods}
\begin{tabular}{|l|l|l|c|} 
\hline
\textbf{Types of analysis} & \textbf{Methodologies} &  \textbf{Input type} \\
\hline
\multirow{6}{*}{Static Analysis} & Symbolic execution & bytecode  \\ 
& Control Flow Graph construction & bytecode \\ 
& Pattern recognition & bytecode \\
& Rule-based analysis & solidity code \\
& Compilation & solidity code \\
& Decompilation & bytecode \\
\hline
\multirow{4}{*}{Dynamic Analysis} & Execution trace at run-time & bytecode \\
& Transaction graph construction & bytecode\\
& Symbolic analysis & bytecode \\
& Validation of true/false positives & bytecode \\
\hline
\multirow{3}{*}{Formal verification} & Using theorem provers & bytecode\\
& Translation of formal language & solidity code \\
& Construction of program logics & bytecode \\
\hline
\end{tabular}\\
\end{table}

\begin{figure*}[!ht]
\centering
\fbox{\includegraphics[width=0.97\textwidth]{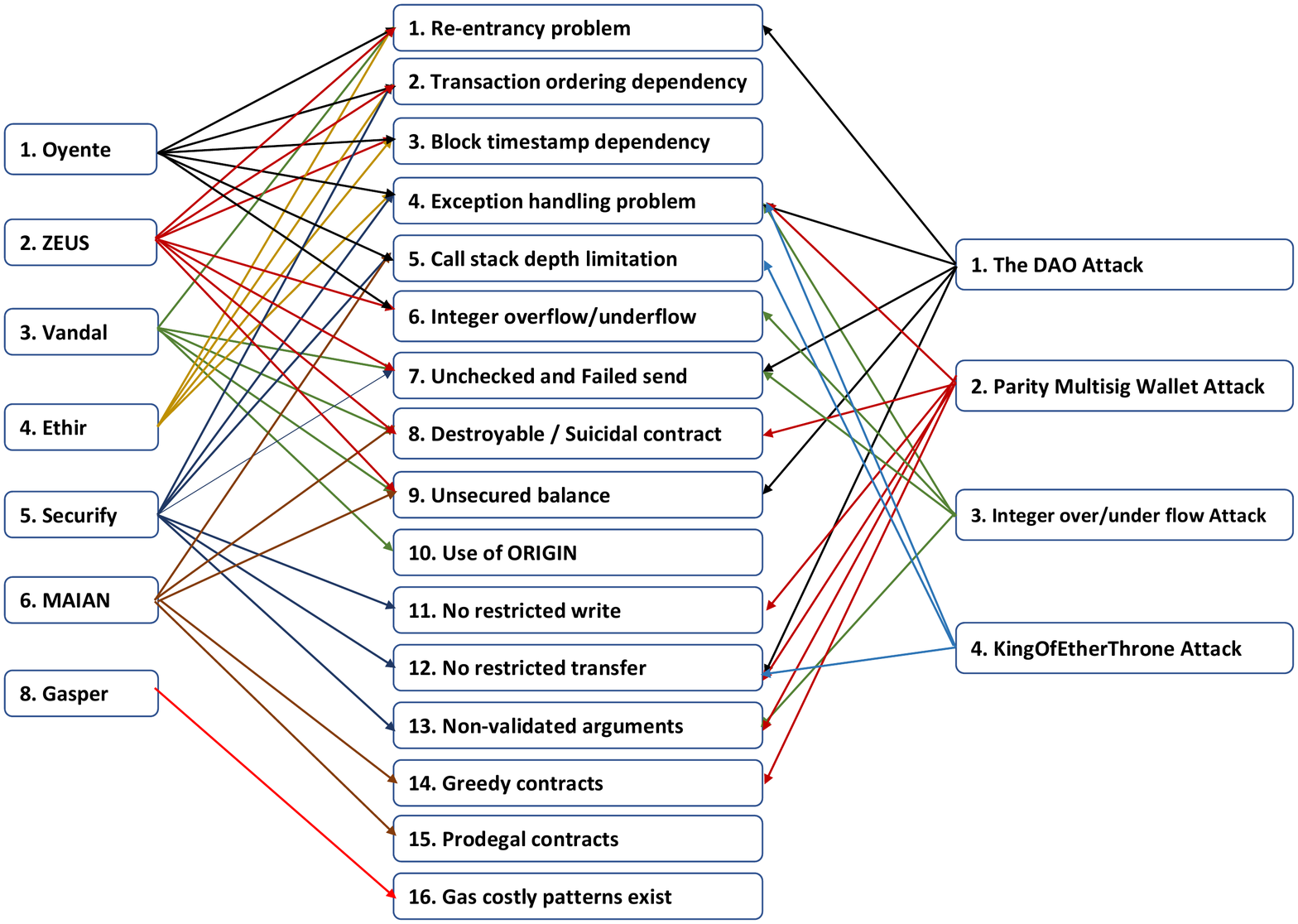}}
\caption{The systematic mapping between Ethereum vulnerabilities, analysis tools, and attacks}
\label{fig:tools_vul_attack_mapping}
\end{figure*}

We categorize the security analysis methods of smart contracts in three types --- static analysis, dynamic analysis, and formal verification methods. Table \ref{tab:analysis_methods} lists the security analysis methods for detecting smart contracts vulnerability using different methodologies and input types. There are several symbolic execution tools to find code vulnerabilities in smart contracts, such as \emph{OYENTE} \cite{RN49}, \emph{MAIAN} \cite{RN86}, \emph{ZEUS} \cite{RN28}, \emph{GASPER} \cite{RN47}, \emph{Securify} \cite{RN116}, \emph{Mythril} \cite{RN208}, and \emph{SmartCheck} \cite{Rn209}. Formal verification methods are high-level analysis on Ethereum bytecodes using theorem provers, such as \emph{isabelle/hol} \cite{RN118}, \emph{KEVM} \cite{RN140}, and \emph{Coq} \cite{RN113, RN119}. This section briefly introduces these analysis methods and compares them with examples. The systematic mapping between identified Ethereum vulnerabilities, detection tools and attacks are presented in Figure \ref{fig:tools_vul_attack_mapping}.

\subsection{Static Analysis}
Static analysis is a way of analyzing a computer program or compiled code in a non run-time environment. The static analysis method inspects the programming code without executing the program. It generally examines all possible code behaviors, vulnerable patterns, and flaws which would be expected in the run-time. This subsection presents a few primary static analysis tools which analyzes the smart contracts security problems and vulnerabilities.

\subsubsection{OYENTE}

Luu et al.~\cite{RN49} investigated the security of the existing smart contracts on the Ethereum network. Several security problems were identified such that  the attackers can manipulate the smart contract execution. Using symbolic execution methods, \emph{OYENTE} is a static analysis tool which detects the security vulnerabilities. The vulnerabilities include transaction ordering dependence, timestamp dependence, mishandled exceptions, and re-entrancy vulnerabilities \cite{RN49}.

\begin{figure}[!ht]
\centering
\fbox{\includegraphics[width=0.47\textwidth]{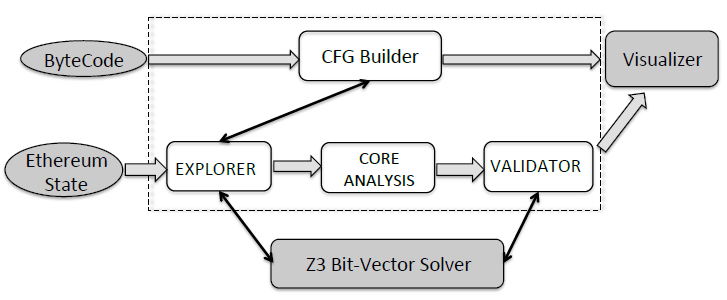}}
\caption{The Architecture of the \emph{OYENTE} Tool \cite{RN49}}
\label{fig:figure22}
\end{figure}

The architecture of the tool \emph{OYENTE} is illustrated in Figure \ref{fig:figure22}. The bytecode of a smart contract and the current global state of Ethereum are taken as inputs. The samples of the smart contracts bytecode are publicly available on the Ethereum network and downloadable via the service named \emph{Etherscan} \cite{RN62}. The initial values of the smart contract variables are extracted from the global state of Ethereum, which improves the accuracy of the analysis. Upon the detection of any problem, \emph{OYENTE} pinpoints the specific line of the smart contract source code which contains any security vulnerability. 

\emph{OYENTE} has four modules \cite{RN49}, namely \texttt{CFGBuilder}, \texttt{Explorer}, \texttt{CoreAnalysis}, and \texttt{Validator}. \texttt{CFGBuilder} builds a control flow graph for the smart contract bytecode. In the control flow graph, each node represents a basic execution block; the edges represent the execution jumps between the blocks. The \texttt{Explorer} executes the smart contract code symbolically. The output from the \texttt{Explorer} are fed as the input to the \texttt{CoreAnalysis} component. The identified vulnerabilities are targeted to implement the logic in the \texttt{CoreAnalysis} module. In the end, the \texttt{Validator} module filters out the false positives from the results, and the final results are visualized to the users. 

\subsubsection{ZEUS}

\emph{ZEUS} \cite{RN28} can verify the correctness of smart contracts and  validate their fairness. Combining an abstract interpreter with a symbolic model checker, \emph{ZEUS} verifies the safe programming practices of the vulnerable smart contracts. According to \cite{RN28}, \emph{ZEUS} outperformed \emph{OYENTE} \cite{RN49} with less false positive rate and less analysis time. The tool \emph{ZEUS} detects six security vulnerabilities in smart contracts including re-entrancy bug, unchecked send, failed send, integer overflow/underflow, block/transaction state dependence and transaction order dependence \cite{RN49,RN50, RN28}.

\emph{ZEUS} consists of three components --- \texttt{policy builder}, \texttt{source code translator}, and \texttt{verifier}. \emph{ZEUS} takes two inputs, that is, the smart contract source code in Solidity and a security policy written in an specific language to verify the vulnerabilities. In the first step, a static analysis is performed to check the smart contract code, while the \texttt{policy builder} inserts the policy predicates as the assert statements at the appropriate places in the source code. The \texttt{source code translator} converts the source code embedded with the policy assertions to LLVM bytecode. Finally, the \texttt{verifier} determines the assertion violations to identify the vulnerable smart contracts. 

\subsubsection*{Formalizing Solidity Semantics} 
\begin{figure}[!ht]
\centering
\includegraphics[width=0.47\textwidth]{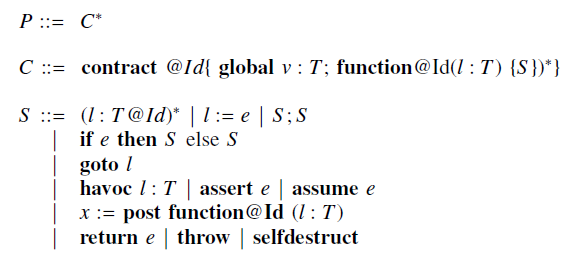}
\caption{An Abstract Language Model for a solidity smart contract\cite{RN28}}
\label{fig:AbsLang}
\end{figure}

An abstract language is defined to capture the related constructs from the Solidity smart contract program \cite{RN28}. Figure \ref{fig:AbsLang} shows the model of the abstract language that is used to formalize the Solidity semantics. A smart contract program consists of a sequence of smart contract declarations. Each smart contract is abstractly implemented with one or more method definitions and logic \cite{RN105, RN127}. The declarations and initialization of methods are stored in the private storage of a contract that is denoted by the keyword \textbf{global}. The variable $Id$ is used to uniquely identify a smart contract. A transaction is the invocation of a publicly accessible contract method. All the methods are defined as a single input variable type of \textbf{$T$}. \textbf{$T$} is a generic variable and can represent collections and struts. There are three types of invocations in Solidity \cite{RN179, RN180, RN182, RN242} internal invocation, external invocation, and call functions. The \textbf{goto} instruction is used to model the internal and external invocations; and the \textbf{post} instruction is used to model the call invocation. The $S$ variable type is defined to represent the body of a contract method. But the \textbf{post} statement can be called with the parameters of smart contracts.

\subsubsection*{Formalizing the Policy Language}
The policy language is formalized for assertion in their abstract language \cite{RN28}. The assertions are used to define the state reachability properties of the smart contract. The policy tuple specification is \texttt{<Sub, Obj, Op, Cond, Res>} which includes the subjects, objects, operations, conditions, and resources \cite{RN210}. The policy tuples are used in \emph{ZEUS} for two reasons: The first reason is to assert the predicate or condition; and the second reason is to extract the correct control location to insert the assert statements into the Solidity source code \cite{RN28}.

\begin{table*}[!ht]
  \centering
  \caption{Tools Versus Detected Vulnerabilities Versus Detected Attacks}
 \label{tab:tools_vulnerabilities_attacks}
\begin{tabular}{ |p{0.1\textwidth}|p{0.6\textwidth}|p{0.2\textwidth}| } 
\hline
\textbf{Tools} & \textbf{Detecting Vulnerabilities} & \textbf{Identified Attacks}\\
\hline
\multirow{1}{12em}{\emph{OYENTE}} & Re-entrancy, Exception handling, Transaction ordering, Block timestamp dependency, Call stack depth limitation & Integer overflow/under flow \emph{The DAO attack} \\ 
\hline
\multirow{1}{12em}{\emph{ZEUS}} & Re-entrancy, Transaction ordering, Block timestamp dependency, Integer over/under flow, unchecked and failed \texttt{send}, Destroyable/Suicidal contract, Unsecured balance &
\emph{The DAO attack}, \emph{Integer Over/Under flow attack}\\
\hline
\multirow{1}{12em}{\emph{Vandal}} & Re-entrancy, Unchecked and failed \emph{send}, Destroyable/Suicidal contract, Unsecured balance, Use of \texttt{Origin} & \emph{The DAO attack}, \emph{Parity multisig wallet attack} \\
\hline
\multirow{1}{12em}{\emph{Ethir}} & Re-entrancy, Exception handling, Transaction ordering, Block timestamp dependency & \emph{The DAO attack} \\
\hline
\multirow{1}{12em}{\emph{Securify}} & Exception handling, Transaction ordering, Call stack depth limitation, Unchecked and Failed \texttt{send}, No Restricted write, No Restricted transfer, Non-validated arguments & \emph{Parity multi sig wallet attack} \\
\hline
\multirow{1}{12em}{\emph{MAIAN}} & Call stack depth limitation, Destroyable/Suicidal contract, Unsecured balance, Greedy contracts and Prodigal contracts  & \emph{Parity Multisig Wallet attack} \\
\hline
\multirow{1}{12em}{\emph{GASPER}} & Gas costly code patterns exist & - \\
\hline
\end{tabular}\\
\end{table*}

\subsubsection{GASPER}
To detect the smart contracts with inefficient gas consumption, a static analysis tool named \emph{GASPER} was developed by Chen et al.~\cite{RN47}. \emph{GASPER} focused on the identification of gas costly patterns from the existing smart contracts. Seven Solidity code patterns were identified in \cite{RN47} which are used by \emph{GASPER} for detection purposes. According to \cite{RN47}, more than 90 percentage of the deployed smart contracts until November 2016, were suffering from some forms of the poorly defined gas cost patterns, and most of these smart contracts consumed a significant amount of gas unnecessarily.

The tool \emph{GASPER} takes smart contract bytecode as the input to identify gas costly patterns. \emph{GASPER} runs symbolic execution on bytecode to find all the reachable code blocks in a candidate smart contract. During the pre-processing step, the \texttt{disasm} command in the Ethereum facilities is used to disassemble the contract bytecode. \emph{GASPER} uses the disassembled results to construct the control flow graph (CFG) of the smart contract. \emph{GASPER} starts a symbolic execution from the root node of the control flow graph and traverses the CFG. Whenever a conditional \texttt{jump} is found during the CFG traversal, \emph{GASPER} checks its feasibility. Specifically, \emph{GASPER} uses the \emph{Z3} solver \cite{RN75} to query the condition whether it is true or false. 

\subsubsection{Vandal}
\emph{Vandal} \cite{RN121} is a security analysis framework for identifying the vulnerabilities in Ethereum smart contracts. An analysis pipeline is used to convert the EVM bytecode to the semantic logic relations. \emph{Vandal} uses the \emph{Souffle} \cite{RN211} language to express the logic specifications for security analysis. \emph{Vandal}'s pipeline has five major components \cite{RN121}: The \texttt{scraper} extracts bytecode of smart contracts in a bulk basis; the \texttt{disassembler} converts the smart contract bytecode into disassemble patterns; the \texttt{decompiler} translates the stack-based bytecode to a register transfer language; on the basis of the register transfer language, the \texttt{extractor} makes logic relations reflecting the program semantics of the smart contract; at last, the \texttt{security analysis} reports any possible vulnerabilities of the examined smart contracts. \emph{Vandal} can identify most of the security vulnerabilities, such as unchecked send, re-entrancy, unsecured balance, destroyable contract, and use of origin problem \cite{RN121, RN50}. 

\subsubsection{Ethir}
\emph{Ethir} \cite{RN119} analyzes Ethereum smart contract bytecode based on the rule-based representations of the control flow graphs (CFG) produced by the \emph{OYENTE} tool \cite{RN49}. \emph{Ethir} produces sound and automated reasoning about the high-level properties of the Ethereum smart contracts. \emph{Ethir} requires \emph{OYENTE} to generate the CFG of EVM code. The first element of \emph{Ethir} is a modified version of \emph{OYENTE} to include all possible jump addresses, since the original \emph{OYENTE} only stores the last value of the jump address \cite{RN49, RN119}. So this modification allows \emph{Ethir} to reconstruct the whole CFG \cite{RN119}. The second element is to translate from EVM bytecode into the rule-based representations by using guarded rules to examine the conditional and unconditional \texttt{jump} instructions. 

\subsubsection{Securify}
\emph{Securify} \cite{RN116} is a fully automated and scalable security analyzer for Ethereum smart contracts. \emph{Securify} checks the smart contract behaviors with respect to a given property, and the result is either safe or unsafe. For finding the violation patterns in the smart contract, \emph{Securify} consists of two components: The dependency graph of each smart contract is symbolically analyzed to extract the semantic information; subsequently, the critical code structure is checked with sufficient conditions to prove whether a property exists or not.

\emph{Securify} checks the important domain-specific properties that are derived from the known attacks, the Solidity recommendations, and the best practices. The security defined specific properties based on the patterns of the known attacks are presented in formal definitions \cite{RN116}. The properties are Ether Liquidity (If a contract has less Ether, it has less Ether liquidity), No writes after the call (There are no writes to the storage variable after any call instructions), Restricted write (Writes to storage is restricted by modifier), Restricted transfer (Ether transfers cannot be invoked by any users who is independent to the senders), Handled exception, Transaction ordering dependency, and Validated arguments (Method parameters should be validated before usage) \cite{RN116}. The \emph{Securify} tool was evaluated with two datasets --- the EVM dataset and the Solidity dataset. The experiment results in \cite{RN116} showed that \emph{Securify} found most of the vulnerabilities and security properties accurately comparing with \emph{OYENTE} \cite{RN49} and \emph{Mythril} \cite{RN208}. 

\subsection{Dynamic Analysis}

Dynamic analysis is a method which checks a programming application while it is executing or in the run-time. It acts similar to an attacker who searches vulnerabilities in a piece of vulnerable code by feeding malicious code or anonymous input to the required functions in a program. Some vulnerabilities would be resulted as false negatives in static analysis, but they can be identified via dynamic analysis method successfully. It also can validate the findings from a static code analyzer.

\subsubsection{MAIAN}
Nikoli\'{c} et al.~\cite{RN34} characterized the smart contract issues as trace vulnerabilities using the detection techniques across a long sequence of invocations of a contract during its run-time. The problematic smart contracts are labeled in three categories --- greedy contracts, prodigal contracts, and suicidal contracts \cite{RN34}. The greedy contracts lock the fund indefinitely while they are alive, and the lock cannot be released in any other conditions. When a smart contract accepts Ether with lack of instructions or unreachable commands, it can become a greedy contract locking the available fund. By default, an Ethereum smart contract returns its funds to the fund owners, when the contract is under attack \cite{RN61, RN98, RN50}. A prodigal contract releases the funds to arbitrary addresses other than to the legitimate owners. Because Ethereum disallows the Ethers held by a smart contract to be released to an arbitrary or unknown address, no actual Ethers will be deposited. An Ethereum smart contract enables a security fallback option of being killed by its owner or by an authorized address \cite{RN34, RN206}. A suicidal contract is vulnerable, because an arbitrary account can kill the contract or force it to execute the \texttt{suicide} instruction \cite{RN34}.

Smart contracts are repeatedly executed during their lifetime \cite{RN73, RN98, RN127, RN145}. A transaction invokes a smart contract and runs a function \cite{RN98}. An execution trace is a sequence of running a contract recorded on the blockchain. \emph{MAIAN} \cite{RN34} considers the execution traces of smart contracts together with the vulnerability categories. An invocation of each run of the contract can exercise an execution path for a given input context. Hence, there may be a chain of effects across a trace of invocations \cite{RN34, RN98}. Considering only one invocation and find a bug on a particular invocation is inefficient. The dynamic analysis tool \emph{MAIAN} uses systematic techniques to find the violations on the defined specific properties of traces in smart contract executions \cite{RN34}.

\begin{figure}[!ht]
\centering
\fbox{\includegraphics[width=0.47\textwidth]{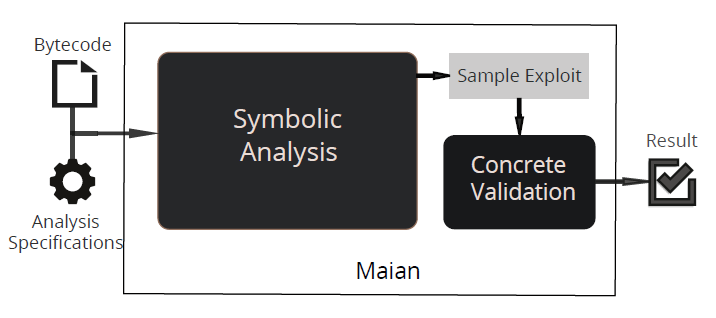}}
\caption{\label{fig:MAIAN} The Architecture of \emph{MAIAN} \cite{RN34}}
\end{figure}

Figure \ref{fig:MAIAN} shows the architecture of \emph{MAIAN}. It has two major components --- \texttt{symbolic analysis} and \texttt{concrete validation}. The contract bytecode and analysis specifications are taken as input to the \texttt{symbolic analysis} component. The analysis specifications contain the vulnerability category and the depth of the search space to define the search operation \cite{RN34}. A custom EVM was implemented to facilitate symbolic execution of smart contract bytecode. The EVM runs for all possible execution traces symbolically for each smart contract candidate. \emph{MAIAN} continues until it reaches a problematic trace with a set of predetermined vulnerability properties. Every execution trace takes a set of symbolic variables as its input. If a contract is detected as vulnerable, then the \texttt{symbolic analysis} component will return concrete values for the specific symbolic variables. The \texttt{concrete validation} component validates the results of the \texttt{symbolic analysis} component. The \texttt{concrete validation} component checks the contract exploitation on a private fork of the Ethereum network \cite{RN34}. It confirms the correctness of bugs found in the candidate smart contract. During the analysis, \emph{MAIAN} does not affect the state of the contract on the main Ethereum blockchain. 

\subsubsection{Graph Construction}

Chen et al.~\cite{RN80} conducted a systematic study on Ethereum by leveraging graph analysis. The major activities on Ethereum were characterized, that is, money transfer, contract creation, and smart contract invocation. The whole internal and external data on Ethereum was collected by modifying Ethereum client using opcodes. New observations and insights were discovered via the construction of three types of graphs \cite{RN80} --- \emph{MFG} (Money Flow Graph), \emph{CCG} (Contract Creation Graph), and \emph{CIG} (Contract Invocation Graph), based on the dynamically collected data. Two new approaches were proposed based on cross-graph analysis to address two security issues in Ethereum. The first application is to find out all accounts controlled by the attacker for a given malicious contract used in digital forensics systems \cite{RN80}; the second application is to detect abnormal contract creation that consumes lots of resources by creating many contracts \cite{RN80}.

\begin{figure}[!ht]
\centering
\fbox{\includegraphics[width=0.47\textwidth]{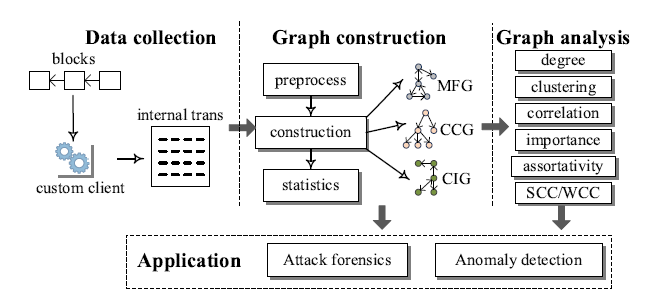}}
\caption{An overview of graph analysis approach  \cite{RN80}}
\label{fig:graph}
\end{figure}

Figure \ref{fig:graph} shows the methodology of the graph analysis approach in \cite{RN80}. The graph-based analysis approach consists of three major phases --- \texttt{data collection}, \texttt{graph construction}, and \texttt{graph analysis}. During \texttt{data collection}, all internal and external transactions data are collected from the Ethereum network. When a contract invokes a method of another smart contract, that is called internal transactions. Since these data are not publicly available in the blockchain, a new approach was introduced to collect internal transactions. The Ethereum client was modified to add instrumentation code using interpretation handler for every EVM opcode. During \texttt{graph construction}, three graphs Money Flow Graph (MFG), Contract Creation Graph (CFG), and Contract Invocation Graph (CIG) are constructed on the basis of all the internal and external transaction data. The transaction data are filtered to exclude the non-relevant transactions in four steps. The relevant transaction data are used to build three types of graphs --- Money Flow Graph (MFG), Contract Creation Graph (CCG), and Contract Invocation Graph (CIG). In a Money Flow Graph (MFG), the edges denote the amount of Ether transferred from one node (account) to another. The sender and the receiver can be an external owned account or a smart contract. A Contract Creation Graph (CCG) captures when a smart contract is created. A Contract Invocation Graph (CIG) is constructed when a transaction executes to call or invoke a smart contract method by an account or from another smart contract. Finally the statistics of the three types of graphs are computed for the \texttt{graph analysis} phase. The graph analysis is conducted on MFG, CCG, and CIG by calculating matrices, such as degree distribution \cite{RN217}, clustering \cite{RN216}, degree correlation \cite{RN215}, node importance \cite{RN217}, Pearson correlation coefficient \cite{RN218}, and strongly/weekly connected component \cite{RN80}. The statistics and matrices provide clear observations and insights \cite{RN80} listed as below. 

\begin{itemize}
\item Most users prefer to transferring money on Ethereum instead of using smart contracts.
\item The smart contracts are not widely used. Many smart contracts are like toy contracts, and lots of them are duplicated.
\item Not all users frequently use the Ethereum network.
\item A small number of developers created lots of smart contracts.
\item The financial applications such as exchange markets, dominate the Ethereum platform.
\end{itemize}

\begin{table*}[!ht]
  \centering
  \caption{Formal Verification Methods and proved properties in smart contracts}
    \label{tab:formal_verification}
\begin{tabular}{ |l|p{0.25\textwidth}|l| } 
\hline
Formal Verification Methods & Proved Properties & Methodologies used \\
\hline
\multirow{2}{12em}{\emph{F*} Framework \cite{RN45}} & run-time safety  &  Solidity translator to \emph{F*}  \\ 
& functional correctness  &  EVM bytecode transator \emph{F*} \\ \hline
\multirow{2}{*}{Formalization using \emph{Isabelle/HOL} \cite{RN118} }& contract correctness & Separation logic and verification conditions\\
& contract termination & Program logic based on execution cost of gas\\
\hline
\multirow{3}{*}{\emph{FEther} using \emph{Coq} \cite{RN113}} & functional correctness & Symbolic execution and higher order logic theorem proofs \\
& Improvement of theorem proving methods of contracts & Verification using Coq \\ 
\hline
\end{tabular}\\
\end{table*}

\subsection{Formal Verification Method}
Formal verification methods use theorem provers or formal methods of mathematics to prove the specific properties in a programming code such as functional correctness, run-time safety, soundness, reliability, and so on. There are a few formal verification analysis conducted to validate and prove vulnerabilities in smart contracts. They used existing theorm provers such as \emph{Coq}, \emph{Isabelle/HOL}, \emph{Lem} and SMT solvers.

\subsubsection{F* Framework}

Bhargavan et al.~\cite{RN45} developed a framework to analyze and verify both run-time safety and the functional correctness of Solidity smart contracts. The Solidity source code and EVM bytecodes are translated to a programming language called \emph{F*}. A language-based approach is developed for verifying smart contracts with the assumptions that the Solidity compiler is not untrustworthy \cite{RN45}, and it is difficult to directly modify EVM due to its intricate semantics and its limited openness \cite{RN87}. 

\begin{figure}[!ht]
\centering
\fbox{\includegraphics[width=0.47\textwidth]{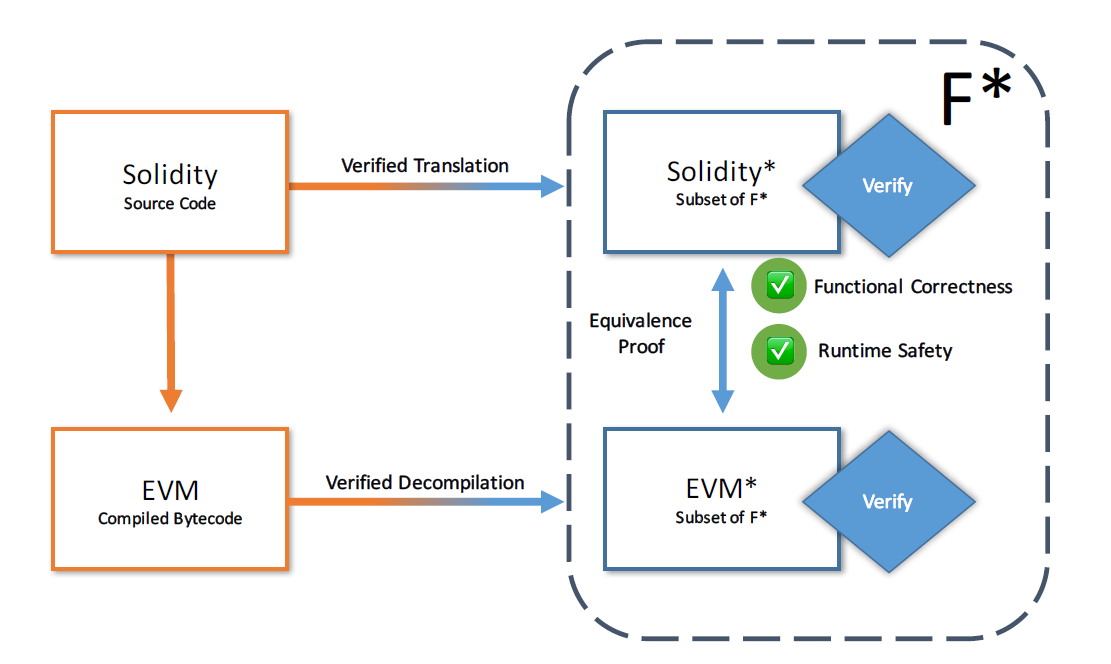} }
\caption{Architecture of the \emph{F*} Framework \cite{RN45}}
\label{fig:F*}
\end{figure}

Figure \ref{fig:F*} shows the architecture of overall framework of \emph{F*} verifier. Two tools are implemented: The first tool is called \emph{Solidity*} which translates the Solidity program to the shallow embedded \emph{F*} programs; the second tool is a decompilier named \emph{EVM*} that converts the EVM bytecode to an equivalent shallow copy of \emph{F*} programs. The source-level functional correctness specificaitions were verified by the \emph{Solidity*} tool for a given piece of Java contract source code. The \emph{EVM*} tool was used to decompile an EVM bytecode of smart contract and analyze the low-level properties, such as gas consumption for each method invocation, execution time, and so on \cite{RN45}. By using both tools, the functional equivalence between the Solidity source code and the EVM bytecode and the correctness of output are verified \cite{RN45}. 

\subsubsection{Formalization using Isabelle/HOL}
Amani et al.~\cite{RN118} built a sound program logic for Ethereum smart contracts bytecode. A proof assistant Isabelle/HOL is used to reason about correctness properties of EVM bytecode based on separation logics \cite{RN118}. All the elements in a program model is carried out by a state. These elements in a state are separated using separation conjunctions as separation logics \cite{RN219}. The formal verification can be used to achieve high-level confidence on the correct behavior of smart contracts. The bytecode sequences were structured into blocks of straight line code and created a program logic for reasoning the behaviors of smart contract code patterns. 

The method of finding correctness properties acts towards of termination based on execution cost of gas in Ethereum. The verification was conducted using a sound program logic at the bytecode level. Smart contract bytecode is divided into two sections as pre-loader and run-time code. Preloader code is used to deploy the contract on Ethereum network. The core functionality of the contract is written in run-time code which are used for verifying smart contracts. Even for a small smart contract, the reasoning about bytecode will have excessively long and repetitive proofs \cite{RN118}. Therefore, it is efficient to the verification conditions using the rules of the logic in \emph{Isabelle} tactics. 

\subsubsection{FEher interpreter using Coq}
FEther is an extensible hybrid verification proof engine that was developed by Yang et al.~\cite{RN113} to improve the theorem proving methods for security of smart contracts. The consistency between smart contracts and its formal model is guaranteed by \emph{FEther} using \emph{Lolisa}. \emph{Lolisa} \cite{RN147} is a formal syntax and semantics for a subset of the solidity programming language. \emph{FEther} combines the symbolic execution with higher order logic theorem proving. A set of automatic strategies in \emph{FEther} helps execute and verify the smart contracts in \emph{Coq}. Its verification process is automated. The segments of verified code is reusable to help verify the specified properties \cite{RN113}. \emph{Coq} is used to interpret and verify the functional correctness in \emph{FEther}. 

\begin{table*}[!ht]
  \centering
  \caption{Analysis tools and open source locations}
    \label{tab:tools_sources}
\begin{tabular}{ |l|l|l| } 
\hline
\textbf{Tool}& \textbf{Source Location} & \textbf{Package Dependencies} \\ 
\hline
\multirow{1}{4em}{\emph{OYENTE}} & \url{https://github.com/melonproject/oyente} &   solc, web3, Z3, Go Ethereum, requests, EVM \\ 
\hline
\multirow{1}{4em}{\emph{MAIAN}} & \url{https://github.com/MAIAN-tool/MAIAN} &   solc, web3, Z3, Go Ethereum, Python, EVM\\ 
\hline
\multirow{1}{4em}{\emph{Securify}} & \url{https://github.com/eth-sri/securify} &   Souffl\'{e}, Java 8, solc, EVM\\
\hline
\multirow{1}{4em}{\emph{Vandal}} & \url{https://github.com/usyd-blockchain/vandal} &   Souffl\'{e}, Python, solc, JSON RPC API, EVM\\ 
\hline
\multirow{1}{4em}{\emph{Ethir}} & \url{https://github.com/costa-group/EthIR} & solc, web3, Z3, Go Ethereum, Python, EVM \\ 
\hline
\multirow{1}{7em}{Graph Analysis} & \url{https://github.com/brokendragon/Ethereum_Graph_Analysis} &   solc, Go Ethereum, Python, EVM\\ 
\hline
\emph{Isabelle/HOL} Proofs & \url{https://github.com/pirapira/eth-isabelle} &   Isabelle2007, Lem Ocaml, Opam packages\\ \hline
\emph{KEVM} framework & \url{https://github.com/kframework/evm-semantics/} &   Pandoc, Java 8 JDK, Opam packages\\ 
\hline
\end{tabular}\\
\end{table*}

\subsection{Comparison between the three analysis Methods}
Here we compare the three analysis methods --- static analysis, dynamic analysis, and formal verification. Both static and dynamic methods use a few similar methodologies such as symbolic execution, transaction/flow graph construction, and validations \cite{RN49, RN50, RN28, RN47, RN128, RN116}. However, static analysis cannot detect vulnerabilities occur during the execution time. In dynamic analysis, the traceability feature is important to identify the erroneous contracts which cause faults in their run-time \cite{RN34}. \emph{MAIAN} traces behind the real execution of smart contracts and finds the vulnerable patterns \cite{RN34}. It would be ensured the reliability of smart contract which passes the test cases throughout the time of its execution or invocations \cite{RN34}. Dynamic analysis tools find a few types of vulnerabilities such as destroyable contract, unsecurred balance, lock and leak contract fund \cite{RN34}. Static analsis tools are able to identify key vulnerable patterns in smart contracts as listed in Table \ref{tab:vulnerabilities_attacks_sins} and \ref{tab:tools_vulnerabilities_attacks}. Formal verification methods are proving specific properties in smart contracts that are performing correct or not. They verify run-time saftey, functional correctness, and sound program logics in smart contracts \cite{RN45,RN118,RN113}. Compare to static and dynamic analysis methods, formal verification methods checks vulnerable patterns using different methodologies, such as separation logic, theorem provers, and translation of EVM byte code to formal languages \cite{RN45,RN118,RN113}.

The static analysis tool \emph{OYENTE} that can detect four major vulnerabilities in smart contracts. The ZEUS tool is able to identify seven vulnerabilities where unchecked send and failed send problems are sub sets of exception handling problem \cite{RN28}. Seven gas costly patterns are defined and identified by the \emph{GASPER} analysis tool \cite{RN47}. The tool \emph{Ethir} used the concept of control flow graph construction from the \emph{OYENTE} tool. \emph{Ethir} is able to find four key vulnerabilities as \emph{OYENTE} detects and includes all possible \texttt{jump} addresses to validate all instructions \cite{RN119}. \emph{Vandal} is detecting five key vulnerabilities using static analysis mechanisms. \emph{Securify} defines seven smart contract vulnerable properties and detects them more accurately \cite{RN116} than \emph{OYENTE} \cite{RN49}. This study categorized \emph{MAIAN} \cite{RN34} as a dynamic analysis tool which defines three errornious contracts and detects them by tracing every invocation paths.

All formal verification methods we discussed \cite{RN45, RN118, RN113} are proving some functional correctness property in smart contracts. They use different methodologies and theorem provers for their verification process as breifed in Table \ref{tab:formal_verification}. They do not detect specific Ethereum vulnerabilities as the analysis tools identify. But they define smart contract correctness and safety properties and able to proof using theorem solving methods. The \emph{F*} framework \cite{RN45} can verify runt-time safety and functional correctness in smart contract execution. 

Comparing the performance between \emph{OYENTE} and \emph{Securify}, it is observed that \emph{OYENTE} \cite{RN49} has missed to report transaction ordering dependency and exception handling problem from few vulnerable contracts \cite{RN116}. Furthermore, \emph{OYENTE} generates more false warnings than \emph{Securify}, when it checks re-entrancy problem in problamatic smart contracts \cite{RN116}.

Only a few tools we analyzed here have published their source codes or executable applications to download as open source. Table \ref{tab:tools_sources} shows the available source links and the required dependencies for each tool.

\section{Research Challenges and Future Directions}
\label{sec:VI}

The DAO attack was occurred due to the two important vulnerablities --- there are an re-entrancy problem and the contract state is updated after sending fund. The re-entrancy problem can be mitigated by using \texttt{address.transfer()} or \texttt{address.send()} functions instead of invoking \texttt{address.call.value()} directly \cite{RN65}. The \texttt{call} function allows caller to make multiple external invocations before the contract state is changed \cite{RN49,RN50}. And developers should aware of updating contract state or balance that should be updated before sending fund to user not after. The tools \emph{OYENTE}, \emph{ZEUS}, \emph{Vandal} and \emph{Ethir} can be used to detect the re-entrancy vulnerability. \emph{Securify} checks the restricted transfer property which help detect the state updating problem and suggest the solution in the relevant line of code \cite{RN116}.

The parity multisig wallet attack happened because of the lack of a proper access modifier to the external library functions \cite{RN66}. The solution for this problem is to use a private modifier to the functions in the external library and use a locking mechanisms to avoid sending fund or changing state without the owner's permission \cite{RN197}. \emph{MAIAN} finds greedy contract that is being frozen and locked its fund indefinitely. This approach will help to find the contracts that call to external functions without having restricted access. The attacks like the partiy multisig wallet problem are partially addressed because it is impossible to avoid all the invocations that are called to the public external functions \cite{RN66}.  

The Integer underflow/overflow attack occurred due to the unchecked send, and the exception handling problem. \emph{ZEUS}, \emph{Vandal}, and \emph{Securify} \cite{RN28, RN116, RN121} are able to detect the unchecked and failed send problem. Further, the latest version of Solidity compiler \cite{RN87} gives warnings to the integer underflow problems while the smart contracts are compiled. Thus this problem is well addressed and able to avoid  many future attacks if the proper version of the Solidity compiler is used \cite{RN116}.

Considering the variety of the key vulnerabilities in Ethereum smart contracts, many vulnerable contracts had already been deployed on the Ethereum blockchain. Because of the immutability feature in smart contracts, the functionalities of deployed smart contracts are unable to modify unless a hard fork. Even though we have analysis tools and verifications methods to detect the buggy contracts \cite{RN49, RN86, RN28, RN121, RN116, RN118, RN119, RN129, RN80, RN47, RN115, RN129, RN130, RN140, RN220}, it is very challenging to eliminate all the vulnerable smart contracts. However, it is recommended to use the Ethereum compiler, analaysis tools, or formal verification methods to test and detect errors before deploy the contracts to the live network. 

The usability of the tools differs significantly. The tools including \emph{OYENTE}, \emph{Securify}, \emph{MAIAN}, and \emph{Vandal} are fully automated analysis tools. The automated tools can be set up easily before analyzing a huge set of smart contracts. \emph{Securify} is a scanning tool available online \cite{RN222} so that smart contract codes can be scanned for possible  vulnerabilities. \emph{OYENTE} provides a docker image \cite{RN223} to deploy the application quickly because a docker image includes all the required dependencies \cite{RN78}. However, only a few formal verification methods have published their source code on github \cite{RN118, RN140}. They are partially automated to verify and prove the correctness properties in smart contracts. The initial setup for formal verification methods takes more time than the symbolic execution tools \cite{RN49, RN86, RN28, RN121, RN116}.

The solidity compiler \texttt{solc} \cite{RN87} is improved well for detecting basic errors and vulnerable patterns in smart contracts during the development phase. Most of the analysis tools depend on the solc compiler to compile smart contract solidity code to bytecode as shown in Table \ref{tab:tools_sources}. As a future work, the detection tools can be integrated with solidity compiler as an external plugin to help the developers identify the vulnerable contracts during the compiling time \cite{RN224, RN225}. Johannes et al.~\cite{RN226} developed an automated tool \emph{teEther} that uses a generic definition of problematic smart contracts to create an exploit for a contract bytecode.

Furthermore, static analysis tools are detecting their specific vulnerabilities as listed in Table \ref{tab:tools_vulnerabilities_attacks}. Seventeen vulnerabilities appeared in the published literature \cite{RN142, RN49, RN50, RN111, RN124}. The logic related problems \cite{RN108} in smart contracts cannot be detected by \emph{OYENTE} \cite{RN49}. It has narrowed down to detect the security bugs relevant to the semantic misunderstandings raised up from smart contracts developers \cite{RN49}. The verification process in \emph{ZEUS} was conducted for the solidity-based smart contracts using an abstract language interpretation approach \cite{RN28}. Kalra et al.~\cite{RN28} demonstrated that \emph{ZEUS} can be extended with a few changes to be compatible to analyze smart contracts on other blockchain platforms \cite{RN28}. \emph{Vandal} framework \cite{RN121} also partly uses an abstract interpretation method, but it analyzes the EVM bytecode directly using its own decompiler for the translation work.

\emph{GASPER} \cite{RN47} can detect seven gas costly patterns in smart contracts. There will be more gas expensive patterns in complex contract programs. Chen et al.~have ensured that they will broad their research on finding more under optimized patterns and detect them by their tool \cite{RN47}. \emph{Ethir} \cite{RN119} framework utilizes the control flow graph methodology developed in \emph{OYENTE} to analyze Ethereum bytecode. But, \emph{Ethir} does not perform any improvement on recovery capability of control flow graph algorithm \cite{RN28}. \emph{Securify} uses Datalog solvers \cite{RN211} to efficiently analyze smart contract code. \emph{Flix} \cite{RN227} enhances the scalablily of analysis process using Datalog. \emph{Securify} \cite{RN116} can utilize these advancements on Datalog solvers as a future development. 

The formal verification methods use different theorem provers such as \emph{Isabelle/HOL}, \emph{F*}, \emph{KEVM}, \emph{Lem}, and \emph{Coq} \cite{RN118, RN140, RN113, RN45}. Since they use  complicated mechanisms, it is not trivial for ordinary users to analyze smart contracts using the formal verification methods. That is, the users must be taught and trained on how the proof method works and on how to read the outputs. Furthermore, the formal verification approach uses a general method to construct code patterns and theorems to prove the security properties of smart contracts using theorem provers \cite{RN118, RN140, RN113, RN45}. Since these provers are semi-automated, the formal verification methods require a significant amount of manual effort to construct the proofs and analysis of smart contracts \cite{RN118, RN121}. Hence, these methods poorly scale for analyzing thousands of smart contracts currently deployed on the Ethereum network \cite{RN62, RN121}. However, the formal verification approach provides accurate and prompt results of validating smart contracts' security, saftey, and soundness properties \cite{RN118, RN130, RN45, RN148, RN149, RN221}.

\section{Conclusion}
\label{sec:VII}
Smart contracts in Ethereum are becoming more applicable as digitalized agent on distributed applications. The security of smart contracts should be ensured to avoid unnecessary losses and malicious attacks. There are several analysis mechanisms implemented to test and assure the correctness and non vulnerable patterns in smart contracts. But developers and users of smart contracts should aware of the accuracy and performance of these analysis methods. Our survey identified the existing vulnerabilities in smart contracts on Ethereum, categorized the security analysis methods in three ways such as static, dynamic, and formal verification. Then we compare the three methods in terms of their performance, coverage of finding vulnerabilities and accuracy. The static and dynamic analysis methods implemented automation tools which are very handy to use and analyse vulnerable contracts. But they detects only their specific defined vulnerable patterns. Formal verification methods uses theorem provers to validate the correctness properties in smart contracts using their interpreted proofs.

\section*{Acknowledgement}
We appreciate the authors who gave permission to reproduce the images from their original papers. We thank to Loi Luu, Antoine Delignat-Lavaud, Ivica Nikoli\'{c} and Yuxiao Zhu for their coordination.

% \nocite{*}
\bibliographystyle{IEEEtran}
\bibliography{export_ref.bib}

\begin{thebibliography}{100}
\providecommand{\url}[1]{#1}
\csname url@samestyle\endcsname
\providecommand{\newblock}{\relax}
\providecommand{\bibinfo}[2]{#2}
\providecommand{\BIBentrySTDinterwordspacing}{\spaceskip=0pt\relax}
\providecommand{\BIBentryALTinterwordstretchfactor}{4}
\providecommand{\BIBentryALTinterwordspacing}{\spaceskip=\fontdimen2\font plus
\BIBentryALTinterwordstretchfactor\fontdimen3\font minus
  \fontdimen4\font\relax}
\providecommand{\BIBforeignlanguage}[2]{{%
\expandafter\ifx\csname l@#1\endcsname\relax
\typeout{** WARNING: IEEEtran.bst: No hyphenation pattern has been}%
\typeout{** loaded for the language `#1'. Using the pattern for}%
\typeout{** the default language instead.}%
\else
\language=\csname l@#1\endcsname
\fi
#2}}
\providecommand{\BIBdecl}{\relax}
\BIBdecl

\bibitem{RN68}
X.~Xu, C.~Pautasso, L.~Zhu, V.~Gramoli, A.~Ponomarev, A.~B. Tran, and S.~Chen,
  ``The blockchain as a software connector,'' in \emph{Software Architecture
  (WICSA), 2016 13th Working IEEE/IFIP Conference on}.\hskip 1em plus 0.5em
  minus 0.4em\relax IEEE, 2016, pp. 182--191.

\bibitem{RN100}
L.~W. Cong and Z.~He, ``Blockchain disruption and smart contracts,'' \emph{The
  Review of Financial Studies}, vol.~32, no.~5, pp. 1754--1797, 2019.

\bibitem{RN152}
O.~Bussmann, ``The future of finance: fintech, tech disruption, and
  orchestrating innovation,'' in \emph{Equity Markets in Transition}.\hskip 1em
  plus 0.5em minus 0.4em\relax Springer, 2017, pp. 473--486.

\bibitem{RN153}
J.~Niehans, ``Transaction costs,'' in \emph{Money}.\hskip 1em plus 0.5em minus
  0.4em\relax Springer, 1989, pp. 320--327.

\bibitem{RN33}
T.~Ahram, A.~Sargolzaei, S.~Sargolzaei, J.~Daniels, and B.~Amaba, ``Blockchain
  technology innovations,'' in \emph{Technology and Engineering Management
  Conference (TEMSCON), 2017 IEEE}.\hskip 1em plus 0.5em minus 0.4em\relax
  IEEE, 2017, Conference Proceedings, pp. 137--141.

\bibitem{RN69}
X.~Xu, I.~Weber, M.~Staples, L.~Zhu, J.~Bosch, L.~Bass, C.~Pautasso, and
  P.~Rimba, ``A taxonomy of blockchain-based systems for architecture design,''
  in \emph{Software Architecture (ICSA), 2017 IEEE International Conference
  on}.\hskip 1em plus 0.5em minus 0.4em\relax IEEE, 2017, pp. 243--252.

\bibitem{RN126}
G.~W. Peters and E.~Panayi, ``Understanding modern banking ledgers through
  blockchain technologies: Future of transaction processing and smart contracts
  on the internet of money,'' in \emph{Banking beyond banks and money}.\hskip
  1em plus 0.5em minus 0.4em\relax Springer, 2016, pp. 239--278.

\bibitem{RN177}
M.~Mainelli and M.~Smith, ``Sharing ledgers for sharing economies: an
  exploration of mutual distributed ledgers (aka blockchain technology),''
  \emph{Journal of Financial Perspectives}, vol.~3, no.~3, 2015.

\bibitem{RN98}
G.~Wood, ``Ethereum: A secure decentralised generalised transaction ledger,''
  \emph{Ethereum project yellow paper}, vol. 151, pp. 1--32, 2014.

\bibitem{RN81}
C.~Cachin, ``Architecture of the hyperledger blockchain fabric,'' in
  \emph{Workshop on Distributed Cryptocurrencies and Consensus Ledgers}, 2016.

\bibitem{RN82}
E.~Androulaki, A.~Barger, V.~Bortnikov, C.~Cachin, K.~Christidis, A.~De~Caro,
  D.~Enyeart, C.~Ferris, G.~Laventman, Y.~Manevich \emph{et~al.}, ``Hyperledger
  fabric: a distributed operating system for permissioned blockchains,'' in
  \emph{Proceedings of the Thirteenth EuroSys Conference}.\hskip 1em plus 0.5em
  minus 0.4em\relax ACM, 2018, p.~30.

\bibitem{RN96}
M.~Raskin and D.~Yermack, ``Digital currencies, decentralized ledgers, and the
  future of central banking,'' National Bureau of Economic Research, Tech.
  Rep., 2016.

\bibitem{RN36}
I.~Weber, V.~Gramoli, A.~Ponomarev, M.~Staples, R.~Holz, A.~B. Tran, and
  P.~Rimba, ``On availability for blockchain-based systems,'' in \emph{Reliable
  Distributed Systems (SRDS), 2017 IEEE 36th Symposium on}.\hskip 1em plus
  0.5em minus 0.4em\relax IEEE, 2017, Conference Proceedings, pp. 64--73.

\bibitem{RN110}
P.~L. Seijas, S.~J. Thompson, and D.~McAdams, ``Scripting smart contracts for
  distributed ledger technology.'' \emph{IACR Cryptology ePrint Archive}, vol.
  2016, p. 1156, 2016.

\bibitem{RN145}
W.~Egbertsen, G.~Hardeman, M.~van~den Hoven, G.~van~der Kolk, and A.~van
  Rijsewijk, ``Replacing paper contracts with ethereum smart contracts,'' 2016.

\bibitem{RN39}
M.~Alharby and A.~van Moorsel, ``Blockchain-based smart contracts: A systematic
  mapping study,'' \emph{arXiv preprint arXiv:1710.06372}, 2017.

\bibitem{RN53}
I.~Eyal, ``Blockchain technology: Transforming libertarian cryptocurrency
  dreams to finance and banking realities,'' \emph{Computer}, vol.~50, no.~9,
  pp. 38--49, 2017.

\bibitem{RN164}
P.~Treleaven, R.~G. Brown, and D.~Yang, ``Blockchain technology in finance,''
  \emph{Computer}, vol.~50, no.~9, pp. 14--17, 2017.

\bibitem{RN154}
S.~A. Abeyratne and R.~P. Monfared, ``Blockchain ready manufacturing supply
  chain using distributed ledger,'' \emph{International Journal of Research in
  Engineering and Technology}, vol.~5, pp. 1--10, 2016.

\bibitem{RN155}
S.~Chen, R.~Shi, Z.~Ren, J.~Yan, Y.~Shi, and J.~Zhang, ``A blockchain-based
  supply chain quality management framework,'' in \emph{2017 IEEE 14th
  International Conference on e-Business Engineering (ICEBE)}.\hskip 1em plus
  0.5em minus 0.4em\relax IEEE, 2017, pp. 172--176.

\bibitem{RN156}
F.~Tian, ``A supply chain traceability system for food safety based on haccp,
  blockchain \& internet of things,'' in \emph{2017 International Conference on
  Service Systems and Service Management}.\hskip 1em plus 0.5em minus
  0.4em\relax IEEE, 2017, pp. 1--6.

\bibitem{RN157}
A.~Azaria, A.~Ekblaw, T.~Vieira, and A.~Lippman, ``Medrec: Using blockchain for
  medical data access and permission management,'' in \emph{2016 2nd
  International Conference on Open and Big Data (OBD)}.\hskip 1em plus 0.5em
  minus 0.4em\relax IEEE, 2016, pp. 25--30.

\bibitem{RN158}
M.~Mettler, ``Blockchain technology in healthcare: The revolution starts
  here,'' in \emph{2016 IEEE 18th International Conference on e-Health
  Networking, Applications and Services (Healthcom)}.\hskip 1em plus 0.5em
  minus 0.4em\relax IEEE, 2016, pp. 1--3.

\bibitem{RN159}
P.~Zhang, D.~C. Schmidt, J.~White, and G.~Lenz, ``Blockchain technology use
  cases in healthcare,'' in \emph{Advances in Computers}.\hskip 1em plus 0.5em
  minus 0.4em\relax Elsevier, 2018, vol. 111, pp. 1--41.

\bibitem{RN160}
K.~N. Griggs, O.~Ossipova, C.~P. Kohlios, A.~N. Baccarini, E.~A. Howson, and
  T.~Hayajneh, ``Healthcare blockchain system using smart contracts for secure
  automated remote patient monitoring,'' \emph{Journal of medical systems},
  vol.~42, no.~7, p. 130, 2018.

\bibitem{RN77}
F.~Knirsch, A.~Unterweger, G.~Eibl, and D.~Engel, ``Privacy-preserving smart
  grid tariff decisions with blockchain-based smart contracts,'' in
  \emph{Sustainable Cloud and Energy Services}.\hskip 1em plus 0.5em minus
  0.4em\relax Springer, 2018, pp. 85--116.

\bibitem{RN161}
E.~Mengelkamp, B.~Notheisen, C.~Beer, D.~Dauer, and C.~Weinhardt, ``A
  blockchain-based smart grid: towards sustainable local energy markets,''
  \emph{Computer Science-Research and Development}, vol.~33, no. 1-2, pp.
  207--214, 2018.

\bibitem{RN162}
C.~Pop, T.~Cioara, M.~Antal, I.~Anghel, I.~Salomie, and M.~Bertoncini,
  ``Blockchain based decentralized management of demand response programs in
  smart energy grids,'' \emph{Sensors}, vol.~18, no.~1, p. 162, 2018.

\bibitem{RN163}
M.~Mylrea and S.~N.~G. Gourisetti, ``Blockchain for smart grid resilience:
  Exchanging distributed energy at speed, scale and security,'' in \emph{2017
  Resilience Week (RWS)}.\hskip 1em plus 0.5em minus 0.4em\relax IEEE, 2017,
  pp. 18--23.

\bibitem{RN125}
K.~Christidis and M.~Devetsikiotis, ``Blockchains and smart contracts for the
  internet of things,'' \emph{Ieee Access}, vol.~4, pp. 2292--2303, 2016.

\bibitem{RN165}
A.~Bahga and V.~K. Madisetti, ``Blockchain platform for industrial internet of
  things,'' \emph{Journal of Software Engineering and Applications}, vol.~9,
  no.~10, p. 533, 2016.

\bibitem{RN166}
N.~Kshetri, ``Can blockchain strengthen the internet of things?'' \emph{IT
  professional}, vol.~19, no.~4, pp. 68--72, 2017.

\bibitem{RN167}
S.~Huh, S.~Cho, and S.~Kim, ``Managing iot devices using blockchain platform,''
  in \emph{2017 19th international conference on advanced communication
  technology (ICACT)}.\hskip 1em plus 0.5em minus 0.4em\relax IEEE, 2017, pp.
  464--467.

\bibitem{RN168}
S.~{\O}lnes, J.~Ubacht, and M.~Janssen, ``Blockchain in government: Benefits
  and implications of distributed ledger technology for information sharing,''
  2017.

\bibitem{RN169}
M.~Staples, S.~Chen, S.~Falamaki, A.~Ponomarev, P.~Rimba, A.~Tran, I.~Weber,
  X.~Xu, and J.~Zhu, ``Risks and opportunities for systems using blockchain and
  smart contracts. data61,'' 2017.

\bibitem{RN31}
C.~Natoli and V.~Gramoli, ``The blockchain anomaly,'' in \emph{Network
  Computing and Applications (NCA), 2016 IEEE 15th International Symposium
  on}.\hskip 1em plus 0.5em minus 0.4em\relax IEEE, 2016, Conference
  Proceedings, pp. 310--317.

\bibitem{RN170}
H.~Kakavand, N.~Kost De~Sevres, and B.~Chilton, ``The blockchain revolution: An
  analysis of regulation and technology related to distributed ledger
  technologies,'' \emph{Bart, The Blockchain Revolution: An Analysis of
  Regulation and Technology Related to Distributed Ledger Technologies (January
  1, 2017)}, 2017.

\bibitem{RN97}
J.~R. Hendrickson, T.~L. Hogan, and W.~J. Luther, ``The political economy of
  bitcoin,'' \emph{Economic Inquiry}, vol.~54, no.~2, pp. 925--939, 2016.

\bibitem{RN171}
P.~Tasca, ``Digital currencies: Principles, trends, opportunities, and risks,''
  \emph{Trends, Opportunities, and Risks (September 7, 2015)}, 2015.

\bibitem{RN131}
M.~Fr{\"o}wis and R.~B{\"o}hme, ``In code we trust?'' in \emph{Data Privacy
  Management, Cryptocurrencies and Blockchain Technology}.\hskip 1em plus 0.5em
  minus 0.4em\relax Springer, 2017, pp. 357--372.

\bibitem{RN70}
M.~Vukoli{\'c}, ``Rethinking permissioned blockchains,'' in \emph{Proceedings
  of the ACM Workshop on Blockchain, Cryptocurrencies and Contracts}.\hskip 1em
  plus 0.5em minus 0.4em\relax ACM, 2017, pp. 3--7.

\bibitem{RN172}
M.~Iansiti and K.~R. Lakhani, ``The truth about blockchain,'' \emph{Harvard
  Business Review}, vol.~95, no.~1, pp. 118--127, 2017.

\bibitem{RN173}
V.~Buterin \emph{et~al.}, ``A next-generation smart contract and decentralized
  application platform,'' \emph{white paper}, 2014.

\bibitem{RN174}
A.~Dubovitskaya, Z.~Xu, S.~Ryu, M.~Schumacher, and F.~Wang, ``Secure and
  trustable electronic medical records sharing using blockchain,'' in
  \emph{AMIA Annual Symposium Proceedings}, vol. 2017.\hskip 1em plus 0.5em
  minus 0.4em\relax American Medical Informatics Association, 2017, p. 650.

\bibitem{RN175}
K.~Jabbar and P.~Bj{\o}rn, ``Infrastructural grind: introducing blockchain
  technology in the shipping domain,'' in \emph{Proceedings of the 2018 ACM
  Conference on Supporting Groupwork}.\hskip 1em plus 0.5em minus 0.4em\relax
  ACM, 2018, pp. 297--308.

\bibitem{RN176}
D.~G. Mamunts, V.~E. Marley, L.~S. Kulakov, E.~M. Pastushok, and A.~V.
  Makshanov, ``The use of authentication technology blockchain platform for the
  marine industry,'' in \emph{2018 IEEE Conference of Russian Young Researchers
  in Electrical and Electronic Engineering (EIConRus)}.\hskip 1em plus 0.5em
  minus 0.4em\relax IEEE, 2018, pp. 69--72.

\bibitem{RN178}
K.~Czachorowski, M.~Solesvik, and Y.~Kondratenko, ``The application of
  blockchain technology in the maritime industry,'' in \emph{Green IT
  Engineering: Social, Business and Industrial Applications}.\hskip 1em plus
  0.5em minus 0.4em\relax Springer, 2019, pp. 561--577.

\bibitem{RN89}
\emph{Blockchain platform: Ethereum}, \url{https://www.ethereum.org/}.

\bibitem{RN92}
\emph{Blockchain platform: EOS}, \url{https://eos.io/}.

\bibitem{RN91}
\emph{Blockchain platform: Lisk}, \url{https://lisk.io/}.

\bibitem{RN90}
\emph{Blockchain platform: Bitcoin}, \url{https://bitcoin.org/en/}.

\bibitem{RN93}
\emph{Blockchain platform: RootStock}, \url{https://www.rsk.co/}.

\bibitem{RN94}
\emph{Blockchain platform: Hyperledger fabric},
  \url{https://www.hyperledger.org/projects/fabric}.

\bibitem{RN61}
\emph{Ethereum Foundation. Ethereum’s white paper}, 2014,
  \url{https://github.com/ethereum/wiki/wiki/White-Paper}.

\bibitem{RN64}
\emph{Summary of Ethereum Upgradeable Smart Contract Research and Development},
  \url{https://blog.indorse.io/ethereum-upgradeable-smart-contract-strategies-456350d0557c}.

\bibitem{RN76}
A.~Unterweger, F.~Knirsch, C.~Leixnering, and D.~Engel, ``Lessons learned from
  implementing a privacy-preserving smart contract in ethereum,'' in \emph{New
  Technologies, Mobility and Security (NTMS), 2018 9th IFIP International
  Conference on}.\hskip 1em plus 0.5em minus 0.4em\relax IEEE, 2018, pp. 1--5.

\bibitem{RN108}
K.~Delmolino, M.~Arnett, A.~Kosba, A.~Miller, and E.~Shi, ``Step by step
  towards creating a safe smart contract: Lessons and insights from a
  cryptocurrency lab,'' in \emph{International Conference on Financial
  Cryptography and Data Security}.\hskip 1em plus 0.5em minus 0.4em\relax
  Springer, 2016, pp. 79--94.

\bibitem{RN179}
M.~Wohrer and U.~Zdun, ``Smart contracts: security patterns in the ethereum
  ecosystem and solidity,'' in \emph{2018 International Workshop on Blockchain
  Oriented Software Engineering (IWBOSE)}.\hskip 1em plus 0.5em minus
  0.4em\relax IEEE, 2018, pp. 2--8.

\bibitem{RN180}
R.~M. Parizi, A.~Dehghantanha \emph{et~al.}, ``Smart contract programming
  languages on blockchains: An empirical evaluation of usability and
  security,'' in \emph{International Conference on Blockchain}.\hskip 1em plus
  0.5em minus 0.4em\relax Springer, 2018, pp. 75--91.

\bibitem{RN142}
G.~Destefanis, M.~Marchesi, M.~Ortu, R.~Tonelli, A.~Bracciali, and R.~Hierons,
  ``Smart contracts vulnerabilities: a call for blockchain software
  engineering?'' in \emph{2018 International Workshop on Blockchain Oriented
  Software Engineering (IWBOSE)}.\hskip 1em plus 0.5em minus 0.4em\relax IEEE,
  2018, pp. 19--25.

\bibitem{RN182}
S.~Wang, Y.~Yuan, X.~Wang, J.~Li, R.~Qin, and F.-Y. Wang, ``An overview of
  smart contract: architecture, applications, and future trends,'' in
  \emph{2018 IEEE Intelligent Vehicles Symposium (IV)}.\hskip 1em plus 0.5em
  minus 0.4em\relax IEEE, 2018, pp. 108--113.

\bibitem{RN49}
L.~Luu, D.-H. Chu, H.~Olickel, P.~Saxena, and A.~Hobor, ``Making smart
  contracts smarter,'' in \emph{Proceedings of the 2016 ACM SIGSAC Conference
  on Computer and Communications Security}.\hskip 1em plus 0.5em minus
  0.4em\relax ACM, 2016, Conference Proceedings, pp. 254--269.

\bibitem{RN86}
\emph{MAIAN: automatic tool for finding trace vulnerabilities in Ethereum smart
  contracts}, \url{https://github.com/MAIAN-tool/MAIAN}.

\bibitem{RN28}
S.~Kalra, S.~Goel, M.~Dhawan, and S.~Sharma, ``Zeus: Analyzing safety of smart
  contracts,'' in \emph{Proceedings of NDSS}, 2018, Conference Proceedings.

\bibitem{RN121}
L.~Brent, A.~Jurisevic, M.~Kong, E.~Liu, F.~Gauthier, V.~Gramoli, R.~Holz, and
  B.~Scholz, ``Vandal: A scalable security analysis framework for smart
  contracts,'' \emph{arXiv preprint arXiv:1809.03981}, 2018.

\bibitem{RN116}
P.~Tsankov, A.~Dan, D.~Drachsler-Cohen, A.~Gervais, F.~Buenzli, and M.~Vechev,
  ``Securify: Practical security analysis of smart contracts,'' in
  \emph{Proceedings of the 2018 ACM SIGSAC Conference on Computer and
  Communications Security}.\hskip 1em plus 0.5em minus 0.4em\relax ACM, 2018,
  pp. 67--82.

\bibitem{RN118}
S.~Amani, M.~B{\'e}gel, M.~Bortin, and M.~Staples, ``Towards verifying ethereum
  smart contract bytecode in isabelle/hol,'' in \emph{Proceedings of the 7th
  ACM SIGPLAN International Conference on Certified Programs and Proofs}.\hskip
  1em plus 0.5em minus 0.4em\relax ACM, 2018, pp. 66--77.

\bibitem{RN119}
E.~Albert, P.~Gordillo, B.~Livshits, A.~Rubio, and I.~Sergey, ``Ethir: A
  framework for high-level analysis of ethereum bytecode,'' in
  \emph{International Symposium on Automated Technology for Verification and
  Analysis}.\hskip 1em plus 0.5em minus 0.4em\relax Springer, 2018, pp.
  513--520.

\bibitem{RN129}
T.~Abdellatif and K.-L. Brousmiche, ``Formal verification of smart contracts
  based on users and blockchain behaviors models,'' in \emph{2018 9th IFIP
  International Conference on New Technologies, Mobility and Security
  (NTMS)}.\hskip 1em plus 0.5em minus 0.4em\relax IEEE, 2018, pp. 1--5.

\bibitem{RN80}
T.~Chen, Y.~Zhu, Z.~Li, J.~Chen, X.~Li, X.~Luo, X.~Lin, and X.~Zhang,
  ``Understanding ethereum via graph analysis,'' in \emph{Proc. INFOCOM}, 2018.

\bibitem{RN47}
T.~Chen, X.~Li, X.~Luo, and X.~Zhang, ``Under-optimized smart contracts devour
  your money,'' in \emph{Software Analysis, Evolution and Reengineering
  (SANER), 2017 IEEE 24th International Conference on}.\hskip 1em plus 0.5em
  minus 0.4em\relax IEEE, 2017, Conference Proceedings, pp. 442--446.

\bibitem{RN115}
S.-M. Lee, S.~Park, and Y.~B. Park, ``Formal specification technique in smart
  contract verification,'' in \emph{2019 International Conference on Platform
  Technology and Service (PlatCon)}.\hskip 1em plus 0.5em minus 0.4em\relax
  IEEE, 2019, pp. 1--4.

\bibitem{RN130}
X.~Bai, Z.~Cheng, Z.~Duan, and K.~Hu, ``Formal modeling and verification of
  smart contracts,'' in \emph{Proceedings of the 2018 7th International
  Conference on Software and Computer Applications}.\hskip 1em plus 0.5em minus
  0.4em\relax ACM, 2018, pp. 322--326.

\bibitem{RN140}
E.~Hildenbrandt, M.~Saxena, N.~Rodrigues, X.~Zhu, P.~Daian, D.~Guth, B.~Moore,
  D.~Park, Y.~Zhang, A.~Stefanescu \emph{et~al.}, ``Kevm: A complete formal
  semantics of the ethereum virtual machine,'' in \emph{2018 IEEE 31st Computer
  Security Foundations Symposium (CSF)}.\hskip 1em plus 0.5em minus 0.4em\relax
  IEEE, 2018, pp. 204--217.

\bibitem{RN220}
S.~Bragagnolo, H.~Rocha, M.~Denker, and S.~Ducasse, ``Smartinspect: solidity
  smart contract inspector,'' in \emph{2018 International Workshop on
  Blockchain Oriented Software Engineering (IWBOSE)}.\hskip 1em plus 0.5em
  minus 0.4em\relax IEEE, 2018, pp. 9--18.

\bibitem{RN50}
N.~Atzei, M.~Bartoletti, and T.~Cimoli, ``A survey of attacks on ethereum smart
  contracts (sok),'' in \emph{International Conference on Principles of
  Security and Trust}.\hskip 1em plus 0.5em minus 0.4em\relax Springer, 2017,
  Conference Proceedings, pp. 164--186.

\bibitem{RN73}
M.~Bartoletti and L.~Pompianu, ``An empirical analysis of smart contracts:
  platforms, applications, and design patterns,'' in \emph{International
  Conference on Financial Cryptography and Data Security}.\hskip 1em plus 0.5em
  minus 0.4em\relax Springer, 2017, pp. 494--509.

\bibitem{RN74}
X.~Li, P.~Jiang, T.~Chen, X.~Luo, and Q.~Wen, ``A survey on the security of
  blockchain systems,'' \emph{Future Generation Computer Systems}, 2017.

\bibitem{RN111}
S.~Rouhani and R.~Deters, ``Security, performance, and applications of smart
  contracts: A systematic survey,'' \emph{IEEE Access}, 2019.

\bibitem{RN120}
I.~Grishchenko, M.~Maffei, and C.~Schneidewind, ``Foundations and tools for the
  static analysis of ethereum smart contracts,'' in \emph{International
  Conference on Computer Aided Verification}.\hskip 1em plus 0.5em minus
  0.4em\relax Springer, 2018, pp. 51--78.

\bibitem{RN117}
------, ``A semantic framework for the security analysis of ethereum smart
  contracts,'' in \emph{International Conference on Principles of Security and
  Trust}.\hskip 1em plus 0.5em minus 0.4em\relax Springer, 2018, pp. 243--269.

\bibitem{RN214}
A.~Mense and M.~Flatscher, ``Security vulnerabilities in ethereum smart
  contracts,'' in \emph{Proceedings of the 20th International Conference on
  Information Integration and Web-based Applications \& Services}.\hskip 1em
  plus 0.5em minus 0.4em\relax ACM, 2018, pp. 375--380.

\bibitem{RN124}
I.-C. Lin and T.-C. Liao, ``A survey of blockchain security issues and
  challenges.'' \emph{IJ Network Security}, vol.~19, no.~5, pp. 653--659, 2017.

\bibitem{RN206}
D.~Harz and W.~Knottenbelt, ``Towards safer smart contracts: A survey of
  languages and verification methods,'' \emph{arXiv preprint arXiv:1809.09805},
  2018.

\bibitem{RN213}
M.~Di~Angelo and G.~Salzer, ``A survey of tools for analyzing ethereum smart
  contracts,'' in \emph{2019 IEEE International Conference on Decentralized
  Applications and Infrastructures (DAPPCON)}.\hskip 1em plus 0.5em minus
  0.4em\relax IEEE, 2019.

\bibitem{RN243}
S.~Lee, C.~Yoon, H.~Kang, Y.~Kim, Y.~Kim, D.~Han, S.~Son, and S.~Shin,
  ``Cybercriminal minds: an investigative study of cryptocurrency abuses in the
  dark web,'' in \emph{Network and Distributed System Security
  Symposium}.\hskip 1em plus 0.5em minus 0.4em\relax Internet Society, 2019,
  pp. 1--15.

\bibitem{RN72}
M.~Di~Angelo and G.~Salzer, ``A survey of tools for analyzing ethereum smart
  contracts,'' in \emph{2019 IEEE International Conference on Decentralized
  Applications and Infrastructures (DAPPCON)}.\hskip 1em plus 0.5em minus
  0.4em\relax IEEE, 2019.

\bibitem{RN65}
\emph{Understanding The DAO Attack}, 2016,
  \url{https://www.coindesk.com/understanding-dao-hack-journalists/}.

\bibitem{RN66}
\emph{An In-Depth Look at the Parity Multisig Bug}, 2016,
  \url{http://hackingdistributed.com/2017/07/22/deep-dive-parity-bug/}.

\bibitem{RN183}
K.~O'hara, ``Smart contracts-dumb idea,'' \emph{IEEE Internet Computing},
  vol.~21, no.~2, pp. 97--101, 2017.

\bibitem{RN45}
K.~Bhargavan, A.~Delignat-Lavaud, C.~Fournet, A.~Gollamudi, G.~Gonthier,
  N.~Kobeissi, N.~Kulatova, A.~Rastogi, T.~Sibut-Pinote, and N.~Swamy, ``Formal
  verification of smart contracts: Short paper,'' in \emph{Proceedings of the
  2016 ACM Workshop on Programming Languages and Analysis for Security}.\hskip
  1em plus 0.5em minus 0.4em\relax ACM, 2016, Conference Proceedings, pp.
  91--96.

\bibitem{RN113}
Z.~Yang and H.~Lei, ``Fether: An extensible definitional interpreter for
  smart-contract verifications in coq.'' \emph{IEEE Access}, 2019.

\bibitem{RN148}
Z.~Yang, H.~Lei, and W.~Qian, ``A hybrid formal verification system in coq for
  ensuring the reliability and security of ethereum-based service smart
  contracts,'' \emph{arXiv preprint arXiv:1902.08726}, 2019.

\bibitem{RN128}
G.~Bigi, A.~Bracciali, G.~Meacci, and E.~Tuosto, ``Validation of decentralised
  smart contracts through game theory and formal methods,'' in
  \emph{Programming Languages with Applications to Biology and Security}.\hskip
  1em plus 0.5em minus 0.4em\relax Springer, 2015, pp. 142--161.

\bibitem{RN149}
\BIBentryALTinterwordspacing
Z.~Yang, ``Formal process virtual machine for smart contracts verification,''
  \emph{International Journal of Performability Engineering}, 2018. [Online].
  Available: \url{http://dx.doi.org/10.23940/ijpe.18.08.p9.17261734}
\BIBentrySTDinterwordspacing

\bibitem{RN221}
S.~K. Lahiri, S.~Chen, Y.~Wang, and I.~Dillig, ``Formal specification and
  verification of smart contracts for azure blockchain,'' \emph{arXiv preprint
  arXiv:1812.08829}, 2018.

\bibitem{RN240}
L.~Alt and C.~Reitwiessner, ``Smt-based verification of solidity smart
  contracts,'' in \emph{International Symposium on Leveraging Applications of
  Formal Methods}.\hskip 1em plus 0.5em minus 0.4em\relax Springer, 2018, pp.
  376--388.

\bibitem{RN242}
R.~M. Parizi, A.~Dehghantanha, K.-K.~R. Choo, and A.~Singh, ``Empirical
  vulnerability analysis of automated smart contracts security testing on
  blockchains,'' in \emph{Proceedings of the 28th Annual International
  Conference on Computer Science and Software Engineering}.\hskip 1em plus
  0.5em minus 0.4em\relax IBM Corp., 2018, pp. 103--113.

\bibitem{RN184}
A.~Baliga, ``Understanding blockchain consensus models,'' in \emph{Persistent},
  2017.

\bibitem{RN185}
I.~Kremenova and M.~Gajdos, ``Decentralized networks: The future internet,''
  \emph{Mobile Networks and Applications}, pp. 1--8, 2019.

\bibitem{RN186}
M.~Valenta and P.~Sandner, ``Comparison of ethereum, hyperledger fabric and
  corda,'' \emph{[ebook] Frankfurt School, Blockchain Center}, 2017.

\bibitem{RN187}
V.~Buterin \emph{et~al.}, ``A next-generation smart contract and decentralized
  application platform,'' \emph{white paper}, 2014.

\bibitem{RN87}
\emph{Solidity source compiler},
  \url{http://solidity.readthedocs.io/en/develop/installing-solidity.html }.

\bibitem{RN188}
C.~Dannen, \emph{Introducing Ethereum and Solidity}.\hskip 1em plus 0.5em minus
  0.4em\relax Springer, 2017.

\bibitem{RN195}
\emph{Ethereum Wallet - MyCrypto}, \url{https://alterdice.com/}.

\bibitem{RN192}
\emph{Ethereum Wallet - MyEtherWallet}, \url{https://www.myetherwallet.com/}.

\bibitem{RN193}
\emph{Ethereum Wallet - MetaMask}, \url{https://metamask.io/}.

\bibitem{RN194}
\emph{Ethereum Wallet - MyCrypto}, \url{https://mycrypto.com/account}.

\bibitem{RN190}
M.~Pusti{\v{s}}ek and A.~Kos, ``Approaches to front-end iot application
  development for the ethereum blockchain,'' \emph{Procedia Computer Science},
  vol. 129, pp. 410--419, 2018.

\bibitem{RN62}
\emph{The Ethereum block explorer}, \url{https://etherscan.io/}.

\bibitem{RN228}
M.~Howard, D.~LeBlanc, and J.~Viega, ``19 deadly sins of software security,''
  \emph{Programming Flaws and How to Fix Them}, 2005.

\bibitem{RN229}
K.~Tsipenyuk, B.~Chess, and G.~McGraw, ``Seven pernicious kingdoms: A taxonomy
  of software security errors,'' \emph{IEEE Security \& Privacy}, vol.~3,
  no.~6, pp. 81--84, 2005.

\bibitem{RN196}
J.~H. Perkins, S.~Kim, S.~Larsen, S.~Amarasinghe, J.~Bachrach, M.~Carbin,
  C.~Pacheco, F.~Sherwood, S.~Sidiroglou, G.~Sullivan \emph{et~al.},
  ``Automatically patching errors in deployed software,'' in \emph{Proceedings
  of the ACM SIGOPS 22nd symposium on Operating systems principles}.\hskip 1em
  plus 0.5em minus 0.4em\relax ACM, 2009, pp. 87--102.

\bibitem{RN109}
B.~Marino and A.~Juels, ``Setting standards for altering and undoing smart
  contracts,'' in \emph{International Symposium on Rules and Rule Markup
  Languages for the Semantic Web}.\hskip 1em plus 0.5em minus 0.4em\relax
  Springer, 2016, pp. 151--166.

\bibitem{RN230}
\emph{Ethereum Classic Network}, \url{https://ethereumclassic.org/}.

\bibitem{RN231}
\emph{The Ethereum Classic 51 Percentage attack is the height of Crypto-Irony},
  \url{https://breakermag.com/the-ethereum-classic-51-attack-is-the-height-of-crypto-irony/}.

\bibitem{RN197}
S.~Palladino, ``The parity wallet hack explained,'' \emph{July-2017.[Online].
  Available: https://blog. zeppelin.
  solutions/on-the-parity-wallet-multisig-hack-405a8c12e8f7}, 2017.

\bibitem{RN198}
H.~Qureshi, ``A hacker stole usd 31 m of ether-how it happened, and what it
  means for ethereum,'' \emph{Appeared at FreeCodeCamp https://medium.
  freecodecamp.
  org/a-hacker-stole-31m-of-ether-how-ithappened-and-what-it-means-for-ethereum-9e5dc29e33ce},
  2017.

\bibitem{RN207}
\emph{Parity Wallet Library}, \url{https://github.com/
  paritytech/parity/blob/4d08e7b0aec46443bf26547b17d10cb302672835/js/src/
  contracts/snippets/enhanced-wallet.sol. }.

\bibitem{RN199}
K.~Iyer and C.~Dannen, ``Contract security,'' in \emph{Building Games with
  Ethereum Smart Contracts}.\hskip 1em plus 0.5em minus 0.4em\relax Springer,
  2018, pp. 91--127.

\bibitem{RN232}
\emph{Ethereum Proposal To “Resurrect” Disabled 360 Mln Dollars Parity
  Contract Shut Down},
  \url{https://cointelegraph.com/news/ethereum-proposal-to-resurrect-disabled-360-mln-parity-contract-shut-down}.

\bibitem{RN233}
\emph{An In-Depth Look at the Parity Multisig Bug},
  \url{http://hackingdistributed.com/2017/07/22/deep-dive-parity-bug/}.

\bibitem{RN200}
\emph{Integer Overflow and Underflow attacks on Smart contracts},
  \url{https://blockgeeks.com/guides/underflow-attacks-smart-contracts/}.

\bibitem{RN201}
\emph{Ethereum Homestead Documentation},
  \url{http://ethdocs.org/en/latest/ether.html}.

\bibitem{RN234}
\emph{Ethereum Known Attacks},
  \url{https://consensys.github.io/smart-contract-best-practices/known_attacks/}.

\bibitem{RN235}
C.~Cowan, C.~Pu, D.~Maier, J.~Walpole, P.~Bakke, S.~Beattie, A.~Grier,
  P.~Wagle, Q.~Zhang, and H.~Hinton, ``Stackguard: Automatic adaptive detection
  and prevention of buffer-overflow attacks.'' in \emph{USENIX Security
  Symposium}, vol.~98.\hskip 1em plus 0.5em minus 0.4em\relax San Antonio, TX,
  1998, pp. 63--78.

\bibitem{RN236}
C.~Cowan, S.~Beattie, J.~Johansen, and P.~Wagle, ``Pointguardtm: Protecting
  pointers from buffer overflow vulnerabilities,'' in \emph{Proceedings of the
  12th conference on USENIX Security Symposium}, vol.~12, 2003, pp. 91--104.

\bibitem{RN237}
P.~Akritidis, M.~Costa, M.~Castro, and S.~Hand, ``Baggy bounds checking: An
  efficient and backwards-compatible defense against out-of-bounds errors.'' in
  \emph{USENIX Security Symposium}, 2009, pp. 51--66.

\bibitem{RN238}
N.~Hasabnis, A.~Misra, and R.~Sekar, ``Light-weight bounds checking,'' in
  \emph{Proceedings of the Tenth International Symposium on Code Generation and
  Optimization}.\hskip 1em plus 0.5em minus 0.4em\relax ACM, 2012, pp.
  135--144.

\bibitem{RN239}
J.~Gao, H.~Liu, C.~Liu, Q.~Li, Z.~Guan, and Z.~Chen, ``Easyflow: Keep ethereum
  away from overflow,'' in \emph{Proceedings of the 41st International
  Conference on Software Engineering: Companion Proceedings}.\hskip 1em plus
  0.5em minus 0.4em\relax IEEE Press, 2019, pp. 23--26.

\bibitem{RN136}
I.~Sergey, A.~Kumar, and A.~Hobor, ``Scilla: a smart contract
  intermediate-level language,'' \emph{arXiv preprint arXiv:1801.00687}, 2018.

\bibitem{RN241}
C.~Liu, H.~Liu, Z.~Cao, Z.~Chen, B.~Chen, and B.~Roscoe, ``Reguard: finding
  reentrancy bugs in smart contracts,'' in \emph{Proceedings of the 40th
  International Conference on Software Engineering: Companion
  Proceeedings}.\hskip 1em plus 0.5em minus 0.4em\relax ACM, 2018, pp. 65--68.

\bibitem{RN202}
L.~W. Cong and Z.~He, ``Blockchain disruption and smart contracts,'' \emph{The
  Review of Financial Studies}, vol.~32, no.~5, pp. 1754--1797, 2019.

\bibitem{RN203}
N.~Grech, M.~Kong, A.~Jurisevic, L.~Brent, B.~Scholz, and Y.~Smaragdakis,
  ``Madmax: Surviving out-of-gas conditions in ethereum smart contracts,''
  \emph{Proceedings of the ACM on Programming Languages}, vol.~2, no. OOPSLA,
  p. 116, 2018.

\bibitem{RN40}
I.~Sergey and A.~Hobor, ``A concurrent perspective on smart contracts,'' in
  \emph{International Conference on Financial Cryptography and Data
  Security}.\hskip 1em plus 0.5em minus 0.4em\relax Springer, 2017, Conference
  Proceedings, pp. 478--493.

\bibitem{RN29}
T.~Dickerson, P.~Gazzillo, M.~Herlihy, and E.~Koskinen, ``Adding concurrency to
  smart contracts,'' in \emph{Proceedings of the ACM Symposium on Principles of
  Distributed Computing}.\hskip 1em plus 0.5em minus 0.4em\relax ACM, 2017,
  Conference Proceedings, pp. 303--312.

\bibitem{RN204}
L.~Yu, W.-T. Tsai, G.~Li, Y.~Yao, C.~Hu, and E.~Deng, ``Smart-contract
  execution with concurrent block building,'' in \emph{2017 IEEE Symposium on
  Service-Oriented System Engineering (SOSE)}.\hskip 1em plus 0.5em minus
  0.4em\relax IEEE, 2017, pp. 160--167.

\bibitem{RN205}
Z.~Gao, L.~Xu, L.~Chen, N.~Shah, Y.~Lu, and W.~Shi, ``Scalable blockchain based
  smart contract execution,'' in \emph{2017 IEEE 23rd International Conference
  on Parallel and Distributed Systems (ICPADS)}.\hskip 1em plus 0.5em minus
  0.4em\relax IEEE, 2017, pp. 352--359.

\bibitem{RN244}
T.~Min and W.~Cai, ``A security case study for blockchain games,'' \emph{arXiv
  preprint arXiv:1906.05538}, 2019.

\bibitem{RN34}
I.~Nikoli\'{c}, A.~Kolluri, I.~Sergey, P.~Saxena, and A.~Hobor, ``Finding the
  greedy, prodigal, and suicidal contracts at scale,'' \emph{arXiv preprint
  arXiv:1802.06038}, 2018.

\bibitem{RN208}
\emph{Mythril - Smart contract security analysis tool},
  \url{https://github.com/ConsenSys/mythril}.

\bibitem{Rn209}
S.~Tikhomirov, E.~Voskresenskaya, I.~Ivanitskiy, R.~Takhaviev, E.~Marchenko,
  and Y.~Alexandrov, ``Smartcheck: Static analysis of ethereum smart
  contracts,'' in \emph{2018 IEEE/ACM 1st International Workshop on Emerging
  Trends in Software Engineering for Blockchain (WETSEB)}.\hskip 1em plus 0.5em
  minus 0.4em\relax IEEE, 2018, pp. 9--16.

\bibitem{RN105}
K.~Lauslahti, J.~Mattila, and T.~Seppala, ``Smart contracts--how will
  blockchain technology affect contractual practices?'' \emph{Etla Reports},
  no.~68, 2017.

\bibitem{RN127}
A.~Mavridou and A.~Laszka, ``Designing secure ethereum smart contracts: A
  finite state machine based approach,'' \emph{arXiv preprint
  arXiv:1711.09327}, 2017.

\bibitem{RN210}
\emph{XACML - eXtensible Access Control Markup Language},
  \url{https://tools.ietf.org/html/rfc7061}.

\bibitem{RN75}
L.~De~Moura and N.~Bj{\o}rner, ``Z3: An efficient smt solver,'' in
  \emph{International conference on Tools and Algorithms for the Construction
  and Analysis of Systems}.\hskip 1em plus 0.5em minus 0.4em\relax Springer,
  2008, pp. 337--340.

\bibitem{RN211}
H.~Jordan, B.~Scholz, and P.~Suboti{\'c}, ``Souffl{\'e}: On synthesis of
  program analyzers,'' in \emph{International Conference on Computer Aided
  Verification}.\hskip 1em plus 0.5em minus 0.4em\relax Springer, 2016, pp.
  422--430.

\bibitem{RN217}
W.~Aiello, F.~Chung, and L.~Lu, ``A random graph model for massive graphs,'' in
  \emph{STOC}, vol. 2000.\hskip 1em plus 0.5em minus 0.4em\relax Citeseer,
  2000, pp. 1--10.

\bibitem{RN216}
M.~E. Newman, ``Random graphs with clustering,'' \emph{Physical review
  letters}, vol. 103, no.~5, p. 058701, 2009.

\bibitem{RN215}
P.~Mahadevan, D.~Krioukov, K.~Fall, and A.~Vahdat, ``Systematic topology
  analysis and generation using degree correlations,'' in \emph{ACM SIGCOMM
  Computer Communication Review}, vol.~36, no.~4.\hskip 1em plus 0.5em minus
  0.4em\relax ACM, 2006, pp. 135--146.

\bibitem{RN218}
G.~Bounova and O.~De~Weck, ``Overview of metrics and their correlation patterns
  for multiple-metric topology analysis on heterogeneous graph ensembles,''
  \emph{Physical Review E}, vol.~85, no.~1, p. 016117, 2012.

\bibitem{RN219}
J.~C. Reynolds, ``Separation logic: A logic for shared mutable data
  structures,'' in \emph{Proceedings 17th Annual IEEE Symposium on Logic in
  Computer Science}.\hskip 1em plus 0.5em minus 0.4em\relax IEEE, 2002, pp.
  55--74.

\bibitem{RN147}
Z.~Yang and H.~Lei, ``Lolisa: Formal syntax and semantics for a subset of the
  solidity programming language,'' \emph{arXiv preprint arXiv:1803.09885},
  2018.

\bibitem{RN222}
\emph{Securify - Security scanner for Ethereum smart contracts},
  \url{https://securify.chainsecurity.com/}.

\bibitem{RN223}
\emph{Securify - Security scanner for Ethereum smart contracts},
  \url{https://hub.docker.com/r/hrishioa/oyente/}.

\bibitem{RN78}
\emph{Git repository - An Analysis Tool for Smart Contracts},
  \url{https://github.com/melonproject/oyente}.

\bibitem{RN224}
A.~Mavridou and A.~Laszka, ``Designing secure ethereum smart contracts: A
  finite state machine based approach,'' \emph{arXiv preprint
  arXiv:1711.09327}, 2017.

\bibitem{RN225}
H.~Rocha, S.~Ducasse, M.~Denker, and J.~Lecerf, ``Solidity parsing using smacc:
  Challenges and irregularities,'' in \emph{Proceedings of the 12th edition of
  the International Workshop on Smalltalk Technologies}.\hskip 1em plus 0.5em
  minus 0.4em\relax ACM, 2017, p.~2.

\bibitem{RN226}
J.~Krupp and C.~Rossow, ``teether: Gnawing at ethereum to automatically exploit
  smart contracts,'' in \emph{27th $\{$USENIX$\}$ Security Symposium
  ($\{$USENIX$\}$ Security 18)}, 2018, pp. 1317--1333.

\bibitem{RN227}
M.~Madsen, M.-H. Yee, and O.~Lhot{\'a}k, ``From datalog to flix: A declarative
  language for fixed points on lattices,'' in \emph{ACM SIGPLAN Notices},
  vol.~51, no.~6.\hskip 1em plus 0.5em minus 0.4em\relax ACM, 2016, pp.
  194--208.

\end{thebibliography}
\end{document}